\documentclass[%
 reprint,
superscriptaddress,
reprint,
 amsmath,amssymb,
 aps,
prb,
floatfix]{revtex4-2}

\usepackage{graphicx}
\usepackage{dcolumn}
\usepackage{bm}
\usepackage{hyperref}
\hypersetup{
    colorlinks=true,
    linkcolor=blue,
    filecolor=blue,      
    urlcolor=blue,
    citecolor=blue
    }

\DeclareMathAlphabet{\mathpzc}{OT1}{pzc}{m}{it}

\usepackage{lipsum}
\usepackage[dvipsnames]{xcolor}
\usepackage{soul}
\newcommand{\p}{\partial}
\newcommand{\hc}{\hat{c}}
\newcommand{\ha}{\hat{a}}

\usepackage{upgreek}

\newcommand{\id} {\mathbb{I}}
\newcommand{\Tr} {{\rm Tr}}

\usepackage{mathtools}
\usepackage{color}

\usepackage{todonotes}

\begin{document}


\title{Gapless Floquet topology}

\author{Gabriel Cardoso}
\affiliation{Tsung-Dao Lee Institute, Shanghai Jiao Tong University, Shanghai, 201210, China}
\author{Hsiu-Chung Yeh}
\affiliation{Center for Quantum Phenomena, Department of Physics,
New York University, 726 Broadway, New York, New York, 10003, USA}
\author{Leonid Korneev}
\affiliation{Department of Physics and Astronomy, Stony Brook University, Stony Brook, NY 11794, USA}
\author{Alexander G. Abanov}
\affiliation{Department of Physics and Astronomy, Stony Brook University, Stony Brook, NY 11794, USA}
\author{Aditi Mitra}
\affiliation{Center for Quantum Phenomena, Department of Physics,
New York University, 726 Broadway, New York, New York, 10003, USA}

\date{\today}

\begin{abstract} 
Symmetry-protected topological (SPT) phases in insulators and superconductors are known for their robust edge modes, linked to bulk invariants through the bulk-boundary correspondence. While this principle traditionally applies to gapped phases, recent advances have extended it to gapless systems, where topological edge states persist even in the absence of a bulk gap. We extend this framework to periodically driven chains with chiral symmetry, revealing the existence of topological edge zero- and $\pi$-modes despite the lack of bulk gaps in the quasienergy spectrum. By examining the half-period decomposition of chiral evolutions, we construct topological invariants that circumvent the need to define the Floquet Hamiltonian, making them more suitable for generalization to the gapless regime. We provide explicit examples, including generalizations of the Kitaev chain and related spin models, where localized $\pi$-modes emerge even when the bulk is gapless at the same quasi-energy as the edge modes. We numerically study the effect of interactions, which give a finite lifetime to the edge modes in the thermodynamic limit with the decay rate consistent with Fermi's Golden Rule.
\end{abstract}

\maketitle


\section{Introduction}

Symmetry-protected topological phases (SPT) in topological insulators and superconductors are typically characterized by their robust gapless edge modes and the principle of bulk-boundary correspondence. This principle links the number of edge modes to the bulk band invariants \cite{hasan2010colloquium,qi2011topological,bernevig2013topological,senthil2015symmetry,wen2017colloquium}. Recently, the concept of bulk-boundary correspondence has been extended to include phases that, despite lacking a bulk gap, still exhibit robust edge modes \cite{kestner2011prediction,fidkowski2011majorana,grover2012quantum,keselman2015gapless,scaffidi2017gapless,verresen2018topology,verresen2020topology,verresen2021gapless,kumar2021multi}. A simple picture of a gapless topological phase is given by the critical phase separating, not a topological from a trivial gapped phase, but two topological gapped phases. A half-integer topological invariant can be defined, and it corresponds to the average of the invariants of the two neighboring gapped phases. It was shown to correspond to the number of topological edge states which survive the closing of the gap, once one discounts the central charge of the bulk system \cite{verresen2018topology}.

Another significant extension of SPT phases is their application to periodically driven, or Floquet, systems. In this case the classification is more complex and requires the introduction of new topological invariants \cite{carpentier2015topological,roy2016abelian,potter2016classification,roy2017periodic,oka2019floquet,harper2020topology,rudner2020band,vu2022dynamic}. For a given symmetry class, the symmetry constraints lead to invariants at each relevant gap in the Floquet spectrum. In the case of chiral-symmetric evolutions, \cite{asboth2013bulk,asboth2014chiral,fruchart2016complex}, the quasienergies $0$ and $\pi$ are singled out for being invariant under chiral symmetry, which leads to a doubling of the invariants. For open boundary conditions, the topological case displays edge-localized zero- and $\pi$-modes whose number is given by appropriate bulk invariants, constituting the Floquet bulk-boundary correspondence.

In this work, we study the fate of the edge zero- and $\pi$-modes when the bulk Floquet spectrum lacks the corresponding gaps. By adapting the generalized notion of topological equivalence of gapless SPT phases to the Floquet case, we find the gapless generalization of the bulk invariants which distinguish between the different classes. In particular, while in most constructions of Floquet invariants it is assumed that the gap is necessary because of the choice of branch cut in defining the Floquet Hamiltonian, our construction shows that this is not essential by avoiding the use of the Floquet Hamiltonian altogether and exploring instead the half-period decomposition discussed in \cite{asboth2014chiral}. We then state and prove the corresponding bulk-boundary correspondence principle relating the bulk invariants to the number of zero- and $\pi$-modes per boundary.

We present simple examples of Floquet drives that realize all possible phases, both with and without gapless modes and bulk gaps at quasienergies $0$ and $\pi$. We make the connection to fermionic Floquet SPTs in the AIII and BDI classes explicit by giving the construction of the fermionic zero- and $\pi$-mode operators. This also connects our discussion to the strong zero- and $\pi$-modes of spin models like the transverse field Ising model and the $XY$ chain \cite{Sen13,Yates19,yates2021strong,yates2022long,mitra2023nonintegrable,yeh2023decay,yeh2023slowly}, where we outline generalizations of these models which display edge modes in the gapless regime. We employ one of these models to numerically study the effect of interactions, which give a finite lifetime to the $\pi$-modes, showing that the decay rate of the edge mode is a polynomial in the interaction strength.

The paper is organized as follows. In Section \ref{sec:chiralfloquet} we define chiral symmetry in Floquet driven systems and review the general formalism. In Section \ref{sec:Fmatrix} we construct the topological bulk invariants and then generalize them to the gapless case in Section \ref{sec:gaplessinvariants}. We prove the bulk-boundary correspondence for these new invariants in Section \ref{sec:bulk-boundary}. In Section \ref{sec:gapbandinv}, we generalize the band invariants to the gapless case, as well as the equations linking gap and band invariants in the Floquet evolutions. We present several examples in Section \ref{sec:examples}, along with their Majorana, complex fermionic, as well as spin forms. 
When employing the spin forms, we also introduce interactions, and present numerical results for the lifetime of the edge modes. We show that in the thermodynamic limit, these edge modes decay due to the perturbation with a lifetime consistent with Fermi's Golden Rule.
In section \ref{sec:conclusions} we summarize our results and discuss future directions. We include in Appendix \ref{app:topinvariants} the proof of the main properties of the topological invariants for arbitrary number of bands. In Appendix \ref{app:bulkboundary} we give additional details on the edge state construction used in the proof of the bulk-boundary correspondence. In Appendix \ref{app:geometryinvariants} we present a geometric interpretation of the band and gap invariants. In Appendix \ref{app:freefermions} we give additional details on the map to fermionic SPTs. In Appendix \ref{app:Numerics} we give additional numerical results for the edge mode dynamics in the presence of interactions.

\begin{figure}[]
    \centering     \includegraphics[width=0.45\textwidth]{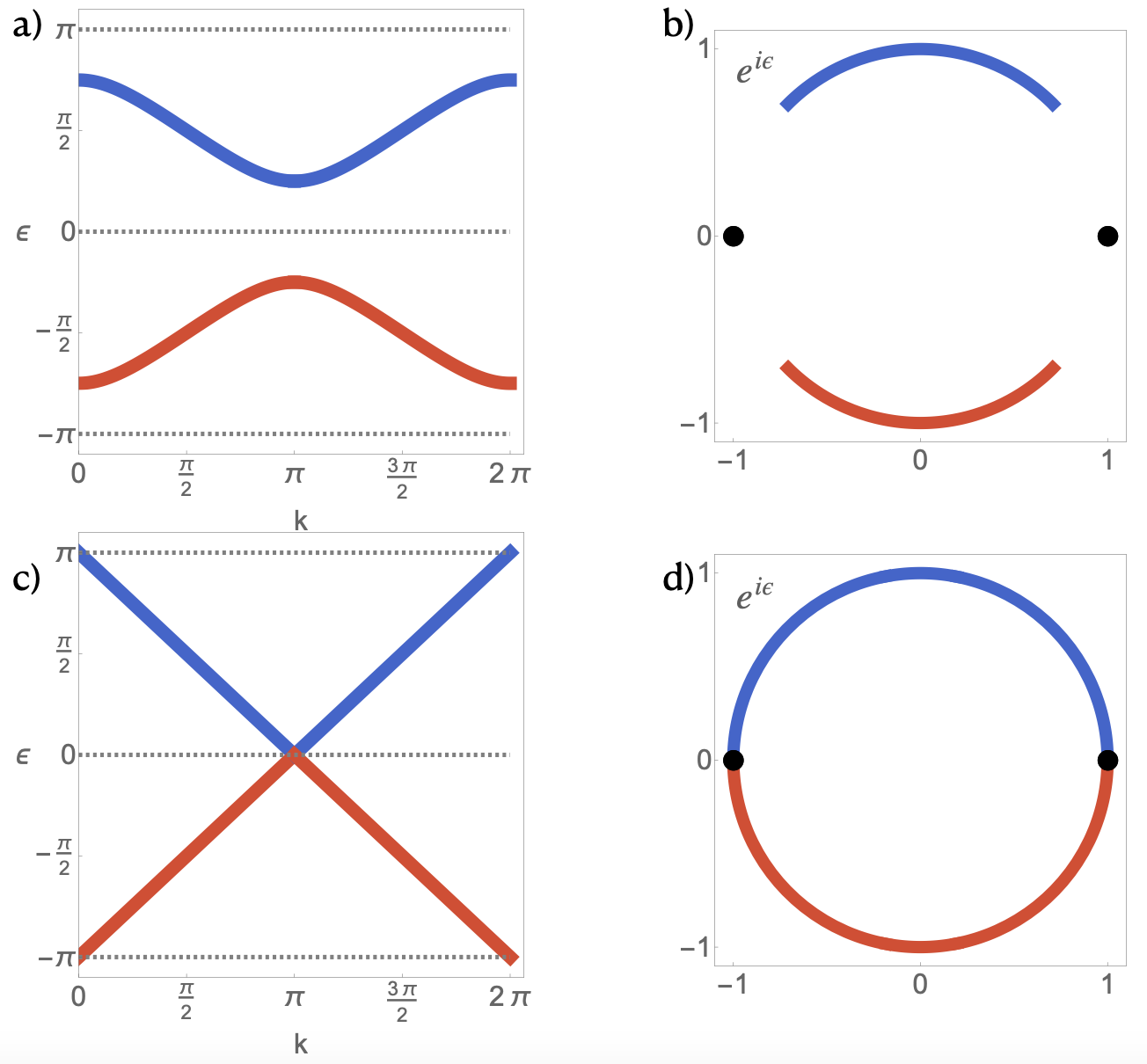}
    \caption{Examples of Floquet spectra with open boundary conditions in a semi-infinite chain (right panels)
    and for periodic boundary conditions (left panels). In a) and b), the bulk is gapped at $\epsilon=0$ and $\epsilon=\pi$, while edge zero- and $\pi$-modes appear in the full spectrum at $e^{i\epsilon}=\pm 1$. This spectrum corresponds to the binary drive \eqref{eq:Fbinary} with $(\alpha_1,\alpha_2)=(2,1)$ in the topological gapped regime $(\vartheta_1,\vartheta_2)=(\frac{3\pi}{2},\frac{5\pi}{4})$. In c) and d), the bulk is gapless at $\epsilon=0$ and $\epsilon=\pi$, but still there are localized edge zero- and $\pi$-modes in the full spectrum with open boundary conditions. We observe this phase for the quaternary drive \eqref{eq:fourstepF}, in the topological gapless regime $(\vartheta,\varphi)=(\frac{3\pi}{2},\pi)$.}
    \label{fig:floquetspectrum}
\end{figure}

\section{Chiral symmetry in Floquet chains}
\label{sec:chiralfloquet}

We start from the general definition of chiral symmetry in a time-dependent system,
\begin{equation}
    \Gamma H(\tau+t)\Gamma = -H(\tau-t),\label{eq:chiralsym}
\end{equation}
where $\tau$ is a base time and the chiral symmetry operator $\Gamma$ is unitary and Hermitian, $\Gamma^2=\id_{N}$, with $N$ the dimension of the single-particle Hilbert space. Chiral symmetry corresponds to sublattice symmetry in the AIII class, while in the BDI class, it arises from the composition of time-reversal and particle-hole symmetries, with this class being even under these symmetries \cite{chiu2016classification,yao2017topological} (see section \ref{sec:fermions}). In a Floquet system, the time-dependence has a discrete translational symmetry,
\begin{equation}
    H(t+T)=H(t).\label{eq:Floquetsym}
\end{equation}
The combination of symmetries (\ref{eq:chiralsym}) and (\ref{eq:Floquetsym}) implies that chiral-symmetric times (CST) come in pairs: if the Hamiltonian has chiral symmetry \eqref{eq:chiralsym} with respect to base time $\tau$, then it also has chiral symmetry with respect to $\tau+T/2$. To simplify the notations, we count time starting from the CST, so that $\tau=0$ and chiral symmetry \eqref{eq:chiralsym} has the simpler form
\begin{equation}
    \Gamma H(t)\Gamma = -H(-t),\label{eq:chiralsym0}
\end{equation}
Additionally, one can define the half-period unitary evolution operator
\begin{equation}
    F=\mathcal{T}\{e^{-i\int_0^{T/2}H(t)dt}\},
    \label{eq:Fdef}
\end{equation}
where $\mathcal{T}$ denotes time ordering. Then it follows that the second half of the dynamics within a period is conjugate to $F$, namely,
\begin{align}
    \mathcal{T}\{e^{-i\int_{T/2}^T H(t)dt}\} = \Gamma F^\dagger \Gamma.\label{eq:conjugate}
\end{align}
This gives a useful decomposition of the Floquet unitaries at the CSTs,
\begin{align}
    U_F(0) &=\mathcal{T}\{e^{-i\int_0^TH(t)dt}\} = \Gamma F^\dagger \Gamma F,\label{eq:UF1}\\
    U_F(T/2) &=\mathcal{T}\{e^{-i\int_{T/2}^{3T/2}H(t)dt}\} = F \Gamma F^\dagger \Gamma,\label{eq:UF2}
\end{align}
where the notation $U_F(t)$ denotes the unitary evolution on a full period starting from time $t$. We will also refer to the Floquet unitaries at CST as
\begin{align}
    &U=U_F(0), &\tilde{U} = U_F(T/2).\label{eq:U12shift}
\end{align}
Note that interchanging $U$ and $\tilde{U}$ corresponds to the half-period shift symmetry $t\mapsto t+T/2$, which acts as
\begin{align}
    &F\mapsto\tilde{F}=\Gamma F^\dagger\Gamma, &U\mapsto\tilde{U}=\Gamma\tilde{F}^\dagger\Gamma\tilde{F}=FUF^\dagger.\label{eq:halfperiodF}
\end{align}
The shift symmetry is an involution, $\tilde{\tilde{U}}=U$, and acts on $U$ as a unitary similarity transformation by $F$. We also note that equations (\ref{eq:UF1},\ref{eq:UF2}) imply that $U,\tilde{U}\in SU(N)$, ie., $\text{Det}[U]=1$, and that they satisfy the chiral symmetry condition
\begin{equation}
    \Gamma U\Gamma = U^\dagger.\label{eq:chiralU}
\end{equation}
We will refer to $U\in SU(N)$ satisfying \eqref{eq:chiralU} as the chiral unitary matrices.

An alternative definition of chiral symmetry is in terms of the Floquet Hamiltonian
\begin{align}
    H_{\text{eff}}=\frac{i}{T}\log_\epsilon U,
\end{align}
Then the chiral symmetry condition \eqref{eq:chiralU} corresponds to
\begin{equation}
    \{H_{\text{eff}},\Gamma\}=0,
\end{equation}
which is identical to the definition of chiral symmetry in the static case. As is well-known, taking the logarithm in the definition of $H_{\rm eff}$ involves the choice of a branch cut at some phase $\epsilon$. In the case of a gapped Floquet spectrum, one can choose the branch cut to lie in the gap, but the definition is less straightforward in the gapless case. In this work, we focus on the construction in terms of the Floquet unitary $U$ and the half-period matrix $F$, which do not require a choice of branch cut.

Note that, for a given chiral Floquet system described by $F$ and $\Gamma$ there are many dynamic realizations in terms of different time-dependent Hamiltonians $H(t)$. It is said that those realizations describe different \emph{micromotions} corresponding to the same stroboscopic dynamics. In the following, our discussion focuses on the stroboscopic properties, and the only aspect of micromotion which we care about is the half-period conjugation of the evolution \eqref{eq:conjugate}, which implies the decompositions (\ref{eq:UF1},\ref{eq:UF2}).

\section{Floquet spectrum and topological invariants}
 \label{sec:Fmatrix}

\begin{figure*}
\centering
\includegraphics[width=1\linewidth]{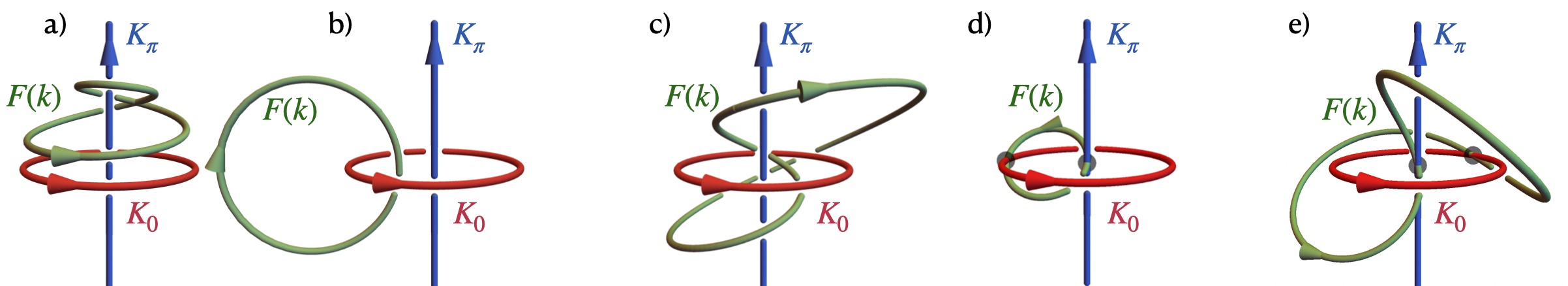}
\caption{Floquet evolutions $F(k)$, seen as oriented loops in three-dimensional space (green). The invariants $\nu_0$, $\nu_\pi$ correspond to the linking numbers of this loop with the oriented loops $K_0$ (red) and $K_\pi$ (blue). a) to c) correspond to the fully gapped case, where $F(k)$ intersects neither $K_0$ nor $K_\pi$. The linking numbers are $(\nu_0,\nu_\pi)=(0,2)$, $(-1,0)$ and $(1,2)$, respectively. In d) and e), the bulk is gapless at both $0$ and $\pi$, which is reflected in the intersections of $F(k)$ and $K_{0,\pi}$ (marked points). However, these two examples are still topologically distinct, with the extended invariants $(\nu_0^+,\nu_0^-,\nu_\pi^+,\nu_\pi^-)=(1,0,1,0)$ (d) and $(2,1,2,1)$ (e), which correspond to no edge modes (d), and one zero- and one $\pi$-mode (e), respectively. (d,e) are obtained from the model in equation \eqref{eq:fourstepF}, with $(\vartheta,\varphi)=(\frac{3\pi}{2},0)$ and $(\frac{3\pi}{2},\pi)$ respectively.}
\label{fig:Fkwinding}
\end{figure*}

The Floquet eigenstates,
\begin{equation}
    U|\Psi\rangle = e^{i\epsilon}|\Psi\rangle,\label{eq:Feigen}
\end{equation}
have a near-translational symmetry in time which follows from the Floquet-Bloch theorem and capture the stroboscopic part of the Floquet dynamics. The phase defines the dimensionless quasienergy $\epsilon$. Unlike in static systems, $\epsilon$ is only defined up to a multiple of $2\pi$. Note that $\tilde{U}$ gives rise to the same Floquet spectrum since it is unitarily related to $U$ by equation \eqref{eq:halfperiodF}. Using the chiral symmetry relation (\ref{eq:chiralU}), it is easy to show that from a given Floquet eigenstate $|\Psi\rangle$ with quasienergy $\epsilon$ one can construct the partner state $\Gamma |\Psi\rangle$ with quasienergy $-\epsilon$. Thus, the quasienergy spectrum is symmetric with respect to the two points $\epsilon=0,\pi$. It follows that one can have unpaired states only at these quasienergies and that these unpaired states cannot be shifted away from $0,\pi$ by any perturbations which preserve chiral symmetry. In this way, the presence of topological bulk bands can lead to edge-localized zero- and $\pi$-modes in the case of open boundary conditions, as shown in Fig.~\ref{fig:floquetspectrum}. Note that in a finite system, the edge states localized on different boundaries can hybridize across the chain, thus shifting away from $0,\pi$, but these energy shifts are exponentially suppressed by the size of the system.

The decomposition \eqref{eq:UF1} relates the Floquet spectrum to the $F$ matrix. First we note that the canonical $U(1)_\chi$ action on $F\mapsto e^{i\chi}F$ acts trivially on the Floquet unitary $U$. In other words, the phase $e^{i\chi}$ only gives information about the micromotion, so that we may consider $F$ as taking values in $SU(N)$. Now note that the condition for $U$ to have a gap at $\epsilon=0$ can be written as
\begin{equation}
    \det(\id_N-U)\neq 0\; \Leftrightarrow\; \det([F,\Gamma])\neq 0,
\end{equation}
and we express this condition as $F\in SU(N)\backslash K_0$, where
\begin{equation}
    K_0 = \{F\in SU(N)|\det([F,\Gamma])=0\},
\end{equation}
denotes the $F$ matrices corresponding to Floquet drives with no gap at $\epsilon=0$. Note that $K_0$ is defined by one complex equation, so it is two real dimensions lower than the total space $SU(N)$. It follows that (one-dimensional) loops can be linked with $K_0$ or, in other words, the space of gapped Floquet evolutions $SU(N)\backslash K_0$ has non-contractible loops, with winding invariant given by
\begin{equation}
    \nu_0 = \frac{1}{i\pi N}\Tr\int [F,\Gamma]^{-1}dF.\label{eq:nutheta}
\end{equation}
In Appendix \ref{app:topinvariants}, we show that the integrand is indeed a closed form counting the winding number associated with a $U(1)$ symmetry of the $F$ parametrization. Here, we note that the gap at $\epsilon=0$ guarantees that the commutator $[F,\Gamma]$ is invertible. As we will discuss below, the topological invariant $\nu_0$ is related to the number of topological zero-modes, and the correspondence can be generalized to the gapless case.

Similarly, the condition that there is a gap at $\epsilon=\pi$ is expressed as
\begin{equation}
    \det(\id_{N}+U)\neq 0\; \Leftrightarrow\; \det(\{F,\Gamma\})\neq 0,
\end{equation}
and we denote it as $F\in SU(N)\backslash K_\pi$, where
\begin{equation}
    K_\pi = \{F\in SU(N)|\det(\{F,\Gamma\})=0\}.
\end{equation}
The space of Floquet evolutions with a gap at $\pi$ also includes a non-contractible loop, with winding given by
\begin{equation}
    \nu_\pi = \frac{1}{i\pi N}\Tr\int \{F,\Gamma\}^{-1}dF,\label{eq:nuphi}
\end{equation}
where this time the gap at $\epsilon=\pi$ guarantees that the anticommutator $\{F,\Gamma\}$ is invertible. We will see that the topological invariant $\nu_\pi$ is related to the number of topological $\pi$-modes, and the correspondence can be generalized to the gapless case.

We remark here that the topological invariants $\nu_{0,\pi}$ have been introduced as invariants of the bulk, that is, in the thermodynamic limit. In the following, we will see that they are related to the number of edge modes in a finite system with open boundary conditions by the bulk-boundary correspondence.

\subsection{Two-band case}

The two-band case $F\in SU(2)$ illustrates the above structure in an intuitive way. Let us take the chiral basis, where $\Gamma=\sigma^z$, and recall that a matrix $F\in SU(2)$ has the standard form
\begin{align}
    F = \begin{pmatrix}
        \zeta & -\eta^*\\
        \eta & \zeta^*
    \end{pmatrix},\;\; |\zeta|^2+|\eta|^2=1,\label{eq:Fcomponents}
\end{align}
where the parametrization in terms of the complex coordinates $(\zeta,\eta)$ identifies $SU(2)$ with the three-sphere $S^3\subset\mathbb{C}^2$. The region $K_0$ where the $\epsilon=0$ gap closes is given by $\eta=0$, which corresponds to the unit circle $(\zeta,\eta)=(e^{i\phi},0)$ parametrized by the phase of $\zeta$. In its complement, direct evaluation of (\ref{eq:nutheta}) gives the invariant
\begin{equation}
    \nu_0 =\frac{1}{2\pi}\int d[\arg(\eta)].\label{eq:nuz}
\end{equation}
Similarly, the region  $K_\pi$ where the $\epsilon=\pi$ gap closes is given by $\zeta=0$, which corresponds to the unit circle $(\zeta,\eta)=(0,e^{i\theta})$ parametrized by the phase of $\eta$. In its complement, direct evaluation of (\ref{eq:nuphi}) gives the invariant
\begin{equation}
    \nu_\pi = \frac{1}{2\pi}\int d[\arg(\zeta)].\label{eq:nuw}
\end{equation}
Equations (\ref{eq:nuz},\ref{eq:nuw}) previously appeared in \cite{asboth2014chiral} as auxiliary formulas in the evaluation of band invariants (see section \ref{sec:gapbandinv}). We see that they are special cases of (\ref{eq:nutheta},\ref{eq:nuphi}), which follow quite generally from chiral symmetry with abritrary number of bands $N$, and do not depend on the choice of basis.

Geometrically, the invariants can be visualized as follows. Consider the infinite bulk system and let us denote by $k\in[0,2\pi)$ the Brillouin zone momentum. Then the Floquet evolution in a 1D chain gives a path $F(k)$ in the space of $SU(2)$ matrices. Using the parametrization $(\zeta(k),\eta(k))$, this can be seen as a closed loop in $S^3$. If there is a gap at $0$, then $\eta(k)$ does not vanish, and the invariant $\nu_0$ \eqref{eq:nuz} gives the linking number of this loop with the loop $K_0$ inside $S^3$. Similarly, $\nu_\pi$ \eqref{eq:nuw} gives the linking number with the loop $K_\pi$ inside $S^3$. To represent these graphically, we can use the stereographic projection to map $S^3$ to $\mathbb{R}^3$,
\begin{equation}
    (\zeta,\eta)\mapsto \frac{1}{1-{\rm Im}(\eta)}\left({\rm Re}(\zeta),{\rm Im}(\zeta),{\rm Re}(\eta)\right),
\end{equation}
so that $F(k)$ is mapped to a loop in three dimensions. $\nu_0$ is the linking number of this loop with the unit circle in the $x$-$y$ plane (corresponding to $K_0$), and $\nu_\pi$ is the linking number with the $z$ axis (corresponding to $K_\pi$, which is seen as a loop closed at infinity). We show some examples in Fig.~\ref{fig:Fkwinding}. The Floquet evolutions $F(k)$ are represented as green loops, while $K_0$ and $K_\pi$ are red and blue loops, respectively. In a) to c), we have the gapped case where $F(k)$ intersects neither $K_0$ nor $K_\pi$. The linking numbers are $(\nu_0,\nu_\pi)=(0,2)$, $(-1,0)$ and $(1,2)$, respectively.

\section{Invariants in the gapless case}
 \label{sec:gaplessinvariants}

The definition of topological invariants $\nu_{0,\pi}$ in the previous section assumed the presence of a gap at the respective quasienergies. For a given bulk Floquet evolution $F(k)$, these cannot change under smooth deformations which do not close the gaps, thus the notion of topological equivalence in the gapped case. This definition of topological equivalence has been applied to both static \cite{chiu2016classification,senthil2015symmetry} and Floquet SPT phases \cite{roy2017periodic,yao2017topological}. In the static case, a generalization to the case of gapless SPTs was proposed \cite{verresen2018topology}, which we extend to the Floquet case. First, we define the pair of central charges $c_{0,\pi}$, where $c_\epsilon$ is given by $1/2$ times the number of times the bulk spectrum crosses quasienergy $\epsilon$, counted with multiplicity. The multiplicity refers to the fact that $c_0$ gets a contribution $1/2\times n$ for a point where the dispersion has a zero of order $n$, $\epsilon\sim \pm(k-k_*)^n$, and affects the entanglement entropy \cite{yates2018central}. Then the notion of topological equivalence in the gapless case is that two Hamiltonians are equivalent if they can be smoothly deformed into one another without changing the central charges $(c_0, c_\pi)$. As we now discuss, $\nu_{0,\pi}$ have simple generalizations to the gapless case which are invariant under this definition of topological equivalence.

Let $(\zeta(k),\eta(k))$ denote the $SU(2)$ parametrization of the half-period evolution $F$ (\ref{eq:Fcomponents}) and suppose that the $\epsilon=0$ gap is closed at the points $\{k_*\}$ of the Brillouin zone. Then the integral
\begin{equation}
    \frac{1}{2\pi}\int_0^{2\pi} dk\p_k[\arg(\eta(k))],\label{eq:nu0k}
\end{equation}
is ill-defined because $\eta(k)=0$ for $k\in\{k_*\}$. Alternatively, one can consider the substitution $z=e^{ik}$, which maps the Brillouin zone to the unit circle in the complex plane, and which defines the analytic continuation $\eta(z)$ away from the unit circle. Then one can recast (\ref{eq:nu0k}) as
\begin{equation}
    \frac{1}{2\pi i}\oint_{S^1(1)}\frac{\eta '}{\eta}dz,\label{nu0zint}
\end{equation}
where we use the notation $S^1(r)$ for the circle of radius $r$ centered at the origin of the complex plane with counterclockwise orientation. In this language, gaplessness corresponds to the presence of $2c_0$ zeros of $\eta(z)$ on the unit circle, at which points the logarithmic derivative in the integrand diverges. However, one can still define the regularized integrals
\begin{align}
    \nu_0^\pm &= \lim_{\delta\to 0}\frac{1}{2\pi i}\oint_{S^1(1\pm\delta)}\frac{\eta '}{\eta}dz\,. \label{eq:nu0pm}
\end{align}
These give two integer invariants, which for generic evolutions are well-defined even in the gapless case.

\begin{figure}[]
    \centering     \includegraphics[width=0.4\textwidth]{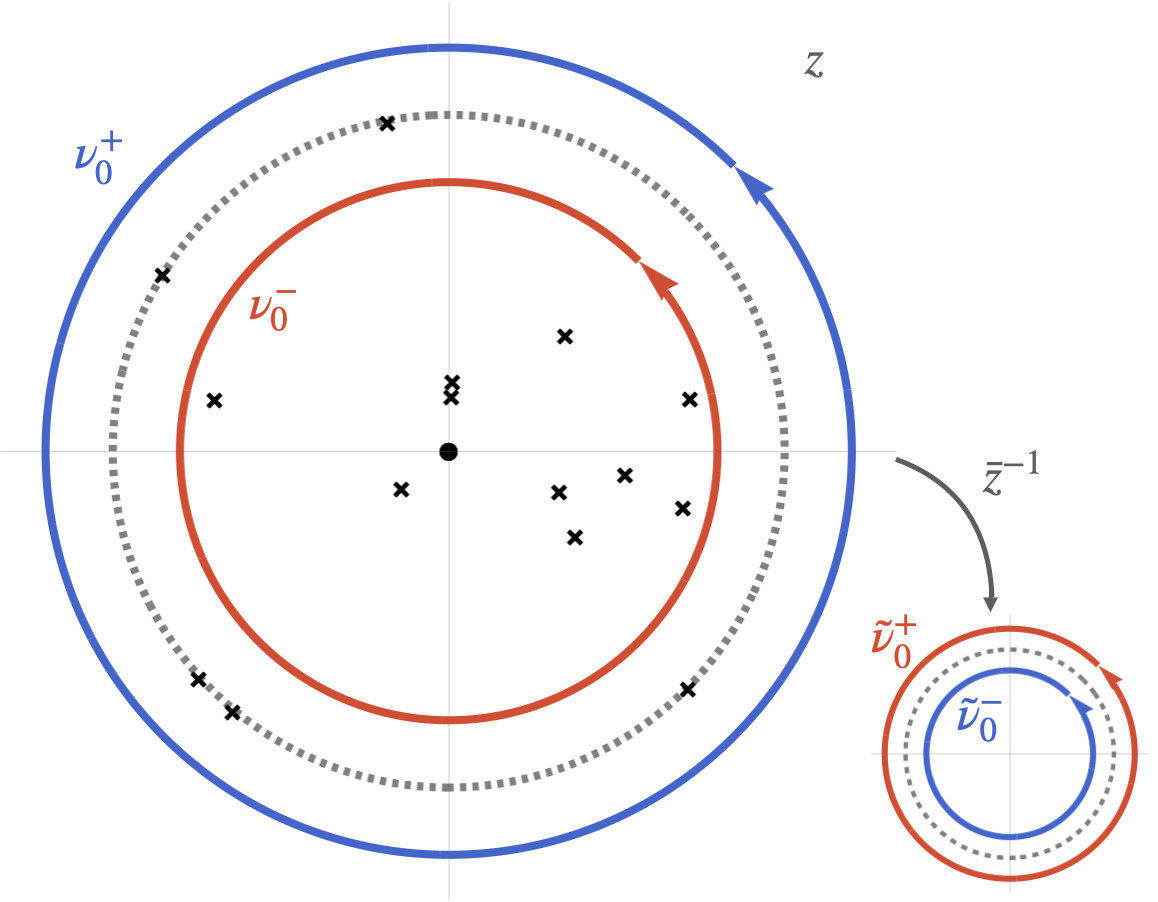}
    \caption{The regularized invariants $\nu_0^\pm$ are defined by deforming the argument integral of $\eta$ to a circle of radius $1\pm\delta$ (blue and red contours). The integrals can be evaluated in terms of the number of zeros (crosses) minus poles (dots) enclosed by the integration path. Thus they differ in the gapless case by the number of zeros on the unit circle (dashed), $2c_0$. The lower inset shows the effect of the Schwarz reflection with respect to the unit circle, equation \eqref{eq:schwarzeta}, on the integration contours.}
    \label{fig:nu0pm}
\end{figure}

In the gapped case, both limits $\nu_0^\pm$ reduce to the original definition of $\nu_0$, equation \eqref{nu0zint}. In the more general case, their difference can be computed as follows. By Cauchy's argument principle, the integral of the logarithmic derivative of a meromorphic function $\eta(z)$ on a counterclockwise-oriented closed contour evaluates to the difference $Z_{\rm in}[\eta]-P_{\rm in}[\eta]$ between the number of zeros and poles in the region enclosed by the contour, counted with multiplicity. Since the quasienergy does not diverge, there are only zeros on the unit circle and no poles, and it follows that (see Fig. \ref{fig:nu0pm})
\begin{equation}
    c_0 = \frac{\nu_0^+-\nu_0^-}{2}.\label{eq:c0nuRL}
\end{equation}
As we show in the next section, both $\nu_0^+$ and $\nu_0^-$ play a role in determining the number of zero-modes per edge in the gapless case.

A similar doubling applies to the invariants for the $\pi$ gap. We define
\begin{align}
    \nu_\pi^\pm &= \lim_{\delta\to 0}\frac{1}{2\pi i}\oint_{S^1(1\pm\delta)}\frac{\zeta '}{\zeta}dz, 
 \label{eq:nupipm}
\end{align}
from which it follows that
\begin{equation}
    c_\pi = \frac{\nu_\pi^+-\nu_\pi^-}{2}.\label{eq:cpinuRL}
\end{equation}

Recall that the invariants $\nu_0$, $\nu_\pi$ can be understood in terms of the linking numbers of the path $F(k)$ with the oriented loops $K_0$, $K_\pi$ in $S^3$. In the gapless case, $F(k)$ intersects these loops, but there is also a generalized notion of a linking number to singular knots that can be applied. It is constructed by taking the average of the two linking numbers obtained when regularizing intersections with opposite orientations \cite{hebda2007linking}. In terms of integral expressions, this definition corresponds to the following half-integer linking numbers 
\begin{align}
    &\frac{\nu_0^-+\nu_0^+}{2}=\frac{1}{2\pi i}\mathcal{P}\oint_{S^1(1)}\frac{\eta '}{\eta}dz,\label{eq:nu0half}\\ 
    &\frac{\nu_\pi^-+\nu_\pi^+}{2}=\frac{1}{2\pi i}\mathcal{P}\oint_{S^1(1)}\frac{\zeta '}{\zeta}dz,\label{eq:nupihalf}
\end{align}
where $\mathcal{P}$ denotes the Cauchy principal value. Generically, expressions (\ref{eq:nu0half},\ref{eq:nupihalf}) are half-integer. They were taken as the topological invariants of gapless Hamiltonians in \cite{verresen2020topology}, where the focus was on the critical gapless phase that lies between the gapped phases with integer invariants $\nu_0$ and $\nu_0+1$. We remark that this definition of the topological invariant should be accompanied by the central charge, so that the different classes are labeled by the pairs $(\nu_0,c_0)$ and $(\nu_\pi,c_\pi)$. Alternatively, the different phases can be  labeled by the double-integer invariants $(\nu_0^+,\nu_0^-)$ and $(\nu_\pi^+,\nu_\pi^-)$, as we do here.

\section{Bulk-boundary correspondence}
 \label{sec:bulk-boundary}

\begin{figure}[]
    \centering     \includegraphics[width=0.45\textwidth]{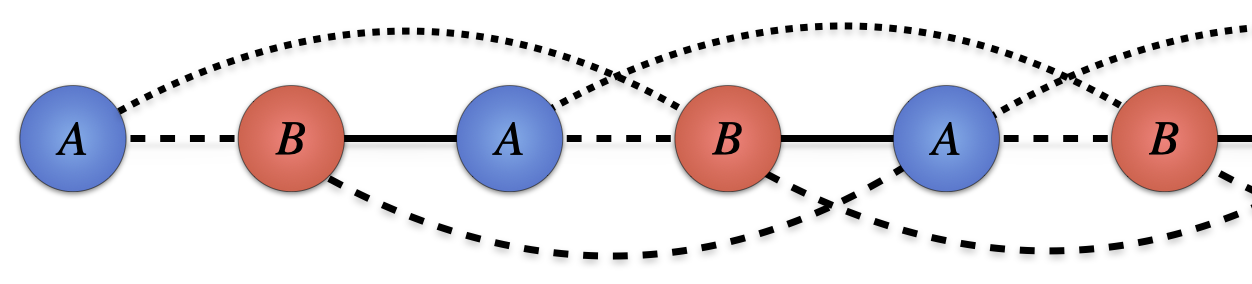}
    \caption{The realization of chiral symmetry as a sublattice symmetry can be visualized on a bi-partite lattice, where the $A$ sites correspond to chirality $\Gamma=+1$ and the $B$ sites correspond to chirality $\Gamma=-1$. In the static case, the chiral symmetry forbids hoppings between sites of equal chirality, such as in the SSH and Kitaev chains (see section \ref{sec:examples}).}
    \label{fig:chirallattice}
\end{figure}

The goal of this section is to relate the invariants computed from the bulk to the number of zero- and $\pi$-modes per boundary. The number of edge states is counted with sign assignments related to the eigenvalues of the chirality operator $\Gamma$. As we saw, chiral symmetry implies that a state $|\Psi\rangle$ of quasienergy $\epsilon$ is accompanied by state $\Gamma |\Psi\rangle$ at quasienergy $-\epsilon$. In the general case where there are many zero- and $\pi$-modes, this means that the action of $\Gamma$ independently preserves the zero-mode subspace and the $\pi$-mode subspace. One can then diagonalize $\Gamma$ in each subspace, so that the zero- and $\pi$-modes come with the additional label of chirality eigenvalue $\pm$. We define the difference between the numbers of positive and negative chirality edge states, on each boundary and at each quasienergy, as
\begin{align}
    &N_0^L = N_0^{L,+}-N_0^{L,-},\label{eq:N0L}\\
    &N_0^R = N_0^{R,-}-N_0^{R,+},\label{eq:N0R}\\
    &N_\pi^L = N_\pi^{L,+}-N_\pi^{L,-},\label{eq:NpiL}\\
    &N_\pi^R = N_\pi^{R,-}-N_\pi^{R,+},\label{eq:NpiR}
\end{align}
where $N_0^{L,+}$ denotes the number of zero-modes of positive chirality localized on the left edge, etc. While the number of states at a fixed chirality depends on details about the boundary conditions, the differences $N_0^L$, $N_0^R$, $N_\pi^L$ and $N_\pi^R$ are topological \footnote{Consider, for example, the change in boundary condition which is equivalent to adding an extra unit cell on the left which is not connected (or weakly connected) to the chain. It will add one extra edge state on the $A$ sublattice (positive chirality) and one extra edge state on the $B$ sublattice (negative chirality), but the difference remains fixed.}, being determined solely by the translation-invariant lattice model in the bulk. We note the difference between the sign assignments at the left and right ends of the chain, as in (\ref{eq:N0L},\ref{eq:N0R}). For example, in the familiar representation of chiral symmetry as a sublattice symmetry, the gapped, static phase with invariant $\nu_0=1$ has one topological left edge state in the $A$ sublattice and one topological right edge state in the $B$ sublattice (Fig. \ref{fig:chirallattice}).

The second subtlety of the above sign assignment is intrinsic to the Floquet case. Let $|\tilde{\Psi}\rangle$, $|\Psi\rangle$ denote a zero- or $\pi$-mode differing by a half-period translation. Then
\begin{align}
    &U|\Psi\rangle = e^{i\epsilon}|\Psi\rangle, &\Gamma|\Psi\rangle = (-1)^{g}|\Psi\rangle,\\
    &\tilde{U}|\tilde{\Psi}\rangle = e^{i\epsilon}|\tilde{\Psi}\rangle, &\Gamma|\tilde{\Psi}\rangle = (-1)^{\tilde{g}}|\tilde{\Psi}\rangle,
\end{align}
where in this notation $g,\tilde{g}\in\{0,1\}$ correspond to the $\pm$ chirality eigenvalues at the chiral symmetric times $\tau=0,T/2$. Using the decomposition (\ref{eq:UF1},\ref{eq:UF2}), we see that
\begin{align}
    e^{i\epsilon}|\tilde{\Psi}\rangle &= \tilde{U}|\tilde{\Psi}\rangle = F\Gamma F^\dagger \Gamma |\tilde{\Psi}\rangle = (-1)^{\tilde{g}}F\Gamma |\Psi\rangle\nonumber \\
    &= (-1)^{g+\tilde{g}}|\tilde{\Psi}\rangle,\label{eq:chiralflip}
\end{align}
where we used that $|\tilde{\Psi}\rangle = F|\Psi\rangle$, since $F$ is the half-period unitary evolution operator. It follows that zero-modes have the same chirality eigenvalue after half period $g=\tilde g$, while $\pi$-modes have the opposite chirality eigenvalue, $g=\tilde g + 1$. In terms of the sublattice symmetry picture, the $\pi$-modes move from one to the other sublattice over half a period. This fact is important because the distinction between different-chirality states appears in the definition of the number of protected edge states (\ref{eq:N0L}-\ref{eq:NpiR}). It implies the transformation rule
\begin{align}
    &\tilde{N}_0=N_0, &\tilde{N}_\pi= -N_\pi,\label{eq:halfperiod}
\end{align}
under the half-period shift. Having defined the invariants and edge state counting, we are ready to state the main result.

\vspace{.2cm}

\noindent\textit{(Bulk-boundary correspondence): In an open chiral Floquet chain, the number of localized zero- and $\pi$-modes per boundary is given by the bulk invariants as
\begin{align}
    &N_0^L=N_0^R=\nu_0^-\theta(\nu_0^-)+\nu_0^+\theta(-\nu_0^+),\label{eq:N0bulkboundary}\\
    &N_\pi^L=N_\pi^R=\nu_\pi^-\theta(\nu_\pi^-)+\nu_\pi^+\theta(-\nu_\pi^+).\label{eq:Npibulkboundary}
\end{align}
These are robust against smooth perturbations to the evolution which do not break chiral symmetry or change the bulk central charges $c_0$ and $c_\pi$.
}

In the gapped case, $\nu_{0,\pi}^+=\nu_{0,\pi}^-$, and these formulas reduce to the usual bulk-boundary correspondence $N_{0,\pi}^L=N_{0,\pi}^R=\nu_{0,\pi}$. In the gapless case, these invariants differ by the central charges and formulas (\ref{eq:N0bulkboundary},\ref{eq:Npibulkboundary}) give the correct generalization. We discuss the proof by outlining how the edge states can be constructed.

\subsection{Edge state construction}

Our strategy for constructing edge states is inspired by the approach used in \cite{verresen2018topology} for static systems, with the difference that in the Floquet case the role of the Hamiltonian is replaced by the half-period evolution $F$. We first recall that, by the argument principle, the winding integrals (\ref{eq:nu0pm},\ref{eq:nupipm}) count the difference between the number of zeros and poles of the functions
\begin{align}
   &\eta(z)=\sum_{m=-\infty}^\infty\eta_m z^m,&\zeta(z)=\sum_{m=-\infty}^\infty\zeta_m z^m ,\label{eq:zetaetaz}
\end{align}
in the region enclosed by the contour. We find that the wavefunctions of the zero- and $\pi$-modes can be constructed from these zeros.

To understand this, let us first consider the translationally invariant chain and write, in the momentum representation,
\begin{equation}
    F = \sum_{k}F(k)|k\rangle\langle k| = \begin{pmatrix}
        \mathcal{Z} &-\mathcal{N}^\dagger\\
        \mathcal{N} &\mathcal{Z}^\dagger
    \end{pmatrix}\,.
 \label{eq:FZH}
\end{equation}
Here the $2\times 2$ matrix form corresponds to the two-dimensional sublattice space of the model. The operators
\begin{align}
    &\mathcal{Z} = \sum_k\zeta(k)|k\rangle\langle k|, &\mathcal{N} = \sum_k\eta(k)|k\rangle\langle k|
\end{align}
are parametrized by complex functions $\zeta(k),\, \eta(k)$ which correspond to \eqref{eq:zetaetaz} at $z=e^{ik}$. They commute,
\begin{align}
    [\mathcal{Z},\mathcal{N}]=0,\quad [\mathcal{Z},\mathcal{N}^\dagger]=0,\quad [\mathcal{Z},\mathcal{Z}^\dagger]=0,\quad [\mathcal{N},\mathcal{N}^\dagger]=0,
 \label{eq:ZHcommute}
\end{align}
and satisfy
\begin{equation}
    \mathcal{Z}^\dagger\mathcal{Z}+\mathcal{N}^\dagger\mathcal{N} =\id.
 \label{eq:ZZHH}
\end{equation}
Their position space representation is given by
\begin{align}
    &\mathcal{Z}=\sum_{m=-\infty}^\infty\zeta_m \hat{T}^{-m}, &\mathcal{N}=\sum_{m=-\infty}^\infty\eta_m \hat{T}^{-m},\label{eq:ZHT}
\end{align}
where the coefficients are the same as in the series expressions \eqref{eq:zetaetaz} and
\begin{equation}
    \hat{T} = \sum_{n}|n+1\rangle\langle n|\label{eq:Tn}
\end{equation}
is the operator of translation by one lattice site. For the finite or half-infinite chain, we choose open boundary conditions such that the $F$ operator is given by the same expressions (\ref{eq:FZH}), (\ref{eq:ZHT}) with $\hat{T}$ truncated to include only the sites of the lattice. We note that while the amplitudes of the edge modes on the first few sites depend on the precise boundary conditions, their decay length into the bulk does not (see Section \ref{sec:examples}).

From the series expressions \eqref{eq:zetaetaz} and \eqref{eq:ZHT}, one can show by direct computation (see Appendix \ref{app:bulkboundary}) that the wavefunction
\begin{eqnarray}
    |\psi_j\rangle = \sum_{n=1}^\infty z_j^{n-1}|n\rangle,\label{eq:smallpsi}
\end{eqnarray}
where $z_j$ is a root of $\zeta(z)$ strickly inside the unit circle, is a solution of
\begin{eqnarray}
    \mathcal{Z}|\psi_j\rangle = 0,\label{eq:Zpsizero}
\end{eqnarray}
with $\mathcal{Z}$ the truncation of \eqref{eq:ZHT} on the half-infinite chain. This means that one can construct a positive-chirality $\pi$-mode
\begin{equation}
    |\Psi_j\rangle = \begin{pmatrix}
        |\psi_j\rangle\\
        0
    \end{pmatrix},
\end{equation}
which is localized on the left boundary since $|z_j|<1$. To see that this is a $\pi$-mode first note that, after half a period, this state evolves to $F|\Psi\rangle$. By the block structure \eqref{eq:FZH}, this state has flipped to negative chirality. Moreover, using equations \eqref{eq:ZHcommute}, \eqref{eq:ZZHH} and \eqref{eq:Zpsizero}, one can show that
\begin{eqnarray}
    U|\Psi_j\rangle = \Gamma F^\dagger\Gamma F |\Psi_j\rangle = -|\Psi_j\rangle.\label{eq:pimodeeqn}
\end{eqnarray}

In the case of $\nu_\pi^->0$, the function $\zeta(z)$ has more zeros than poles inside the unit disk. We note that only the difference is topological, since poles and zeros can cancel each other out by moving them inside the disk (which does not change the central charge of the bulk). For each of the $\nu_\pi^-$ extra zeros, one can follow the above procedure to construct a positive-chirality $\pi$-mode on the left, so that
\begin{equation}
    N_\pi^L = \nu_\pi^-.\label{eq:NpinupiL}
\end{equation}
Finally, note that since $\nu_\pi^+\geq\nu_\pi^-$, equation \eqref{eq:Npibulkboundary} is satisfied, with the second contribution vanishing.

The second contribution plays a role when there are more poles than zeros inside the unit disk, since poles contribute negatively to the invariants. In this case one can construct negative chirality states as follows. Let $|\psi_j\rangle$ be given by equation \eqref{eq:smallpsi}, where $z_j$ is a root of the function
\begin{equation}
    \tilde{\zeta}(z)=\overline{\zeta(\bar{z}^{-1})}\label{eq:schwarzeta},
\end{equation}
strictly inside the unit disk. Then one can show that
\begin{eqnarray}
    \mathcal{Z}^\dagger|\psi_j\rangle = 0.\label{Zdagpsizero}
\end{eqnarray}
Defining the negative-chirality state
\begin{equation}
    |\Psi_j\rangle = \begin{pmatrix}
        0\\
        |\psi_j\rangle
    \end{pmatrix},
\end{equation}
one can then check that it is a $\pi$-mode by evaluating equation \eqref{eq:pimodeeqn} as above.

The transformation \eqref{eq:schwarzeta} is known as the Schwarz reflection of the function $\zeta$ with respect to the unit circle \cite{lang2013complex}. By the previous arguments, we see that if $\tilde{\zeta}$ has winding $\tilde{\nu}_\pi^->0$, then we can construct this many negative chirality $\pi$-modes. Finally, we note that
\begin{equation}
    \tilde{\nu}_\pi^- = -\nu_\pi^+,
\end{equation}
since the Schwarz reflection exchanges the inner and outer regions of the unit circle (see figure \ref{fig:nu0pm}). Alternatively, an excess of poles of $\zeta$ inside the unit disk is accompanied by an excess of zeros of $\tilde{\zeta}$ inside the unit disk. Because negative-chirality states are counted with a minus sign, in this case we get
\begin{equation}
    N_\pi=-\tilde{\nu}_\pi^-=\nu_\pi^+,
\end{equation}
and equation \eqref{eq:Npibulkboundary} is again satisfied, since in this case the first term vanishes (because $\nu_\pi^-\leq\nu_\pi^+<0$).

We also note that the fact that $\nu_\pi^+$ counts the number of $\pi$-modes instead of $\nu_\pi^-$ in the case of negative winding, guarantees consistency with the half-period chirality flip \eqref{eq:halfperiod}. To understand this, first note that from equations (\ref{eq:halfperiodF},\ref{eq:Fcomponents}), the half-period shift corresponds to
\begin{align}
    (\tilde{\zeta}(k),\tilde{\eta}(k))=(\zeta^*(k),\eta(k)).
\end{align}
In terms of its analytic extension, the transformation of $\zeta$ again corresponds to the Schwarz reflection formula \eqref{eq:schwarzeta}, so that under the half-period shift the windings change by
\begin{align}
    (\tilde{\nu}_\pi^+,\tilde{\nu}_\pi^-)= -(\nu_\pi^-,\nu_\pi^+).\label{eq:halfperiodnu}
\end{align}
Substituting in \eqref{eq:Npibulkboundary}, one finds that the full expression satisfies $\tilde{N}_\pi=-N_\pi$, even in the gapless case. 

Interestingly, it is possible for $\nu_\pi^-$ to be negative but for $\nu_\pi^+$ to be positive, in which case there are no $\pi$-modes of either chirality. To understand this in the context of the previous discussion, let us summarize the condition to have $\pi$-modes. These occur when  either $\zeta(z)$ or $\tilde{\zeta}(z)$ have more zeros than poles strictly inside the unit disk. The number of zeros minus poles inside the unit disk is $\nu_\pi^-$ and $-\nu_\pi^+$  for $\zeta(z)$ and $\tilde{\zeta}(z)$ respectively, due to the Schwarz reflection. Thus the existence of the $\pi$-mode requires $\nu_\pi^- > 0$ or $-\nu_\pi^+>0$. But we also know that $\nu_\pi^- \leq \nu_\pi^+$ (where the equality applies to the gapped case). Therefore, only the cases $\nu_\pi^- \leq \nu_\pi^+ < 0$ and $0 < \nu_\pi^- \leq \nu_\pi^+$ allow $\pi$-modes, while the case $\nu_\pi^- < 0 < \nu_\pi^+$ does not give any $\pi$ modes. 

One can apply the same considerations to construct zero modes from the zeros of $\eta(z)$ and $\tilde{\eta}(z)$. Then equations \eqref{eq:Zpsizero} and \eqref{Zdagpsizero} are replaced by $\mathcal{N}|\psi\rangle =0$ and $\mathcal{N}^\dagger|\psi\rangle=0$, respectively. Looking back at the block structure of $F$ in \eqref{eq:FZH}, we see that the zero-modes have the same chirality after half a period, as expected. 

To address the correspondence for right edge states in the case of a finite chain, we consider the reflection of the chain. It acts on the lattice sites as $|n\rangle\mapsto |L-n\rangle$ and on the chirality space as $(u, v)^T\mapsto (v, u)^T$. Then, from our definitions (\ref{eq:N0L}-\ref{eq:NpiR}), it acts on the number of edge states as
\begin{align}
    &N_0^L\leftrightarrow N_0^R, &N_\pi^L\leftrightarrow N_\pi^R.
\end{align}
On the other hand, one can check that the windings $(\nu_0^+,\nu_0^-,\nu_\pi^+,\nu_\pi^-)$ are all invariant under spatial inversion, so that the bulk-boundary correspondence on the right boundary follows from the result on the left boundary.

\section{Gap and band invariants}\label{sec:gapbandinv}

In Floquet phases, we distinguish between gap and band invariants. The first are directly related to the number of edge states and can be computed from the $F$ operator using our construction above. The band invariants, on the other hand, were explored earlier in the literature \cite{asboth2013bulk,asboth2014chiral}. They can be computed from the Floquet Hamiltonian and capture the topology of Floquet Bloch eigenfunctions. We discuss the connection between band and gap invariants using the chiral decomposition of the Floquet unitary.

First note that, if the Floquet spectrum has gaps at both $0$ and $\pi$ quasienergies, then the square of the Floquet unitary $U^2$ does not contain eigenvalue $1=e^{i0}=e^{i2\pi}$ in its spectrum. On the other hand, using the chirality condition \eqref{eq:chiralU} we have $\id-U^2=[\Gamma,U]\Gamma U$. Then the gap condition,
\begin{equation}
    \det(\id-U^2)\neq 0\; \Leftrightarrow\; \det([U,\Gamma])\neq 0,
\end{equation}
is translated to the invertibility of the commutator $[U,\Gamma]$. One can then define the band invariants as
\begin{align}
    &\omega=\frac{1}{i\pi N}\Tr\int [U,\Gamma]^{-1}dU,\label{eq:omega1U}\\
    &\tilde{\omega}=\frac{1}{i\pi N}\Tr\int [\tilde{U},\Gamma]^{-1}d\tilde{U}.\label{eq:omega2U}
\end{align}
These are related to the gap invariants $\nu_{0,\pi}$. For simplicity, consider the two-band case in the chiral basis, where the Floquet unitary has the form
\begin{align}
    &U=\begin{pmatrix}
        a& -q^*\\
        q & a
    \end{pmatrix} &a^2+|q|^2=1.\label{eq:UFcxform}
\end{align}
The diagonal entry $a$ is real due to the chirality condition (\ref{eq:chiralU}). In terms of these components,
\begin{align}
    &\omega=\frac{1}{2\pi}\int d[\arg(q)], &\tilde{\omega}=\frac{1}{2\pi}\int d[\arg(\tilde{q})].
\end{align}
Note that, although \eqref{eq:halfperiodF} implies that the quasienergy spectra of $U$ and $\tilde{U}$ are identical, the windings can still be different, leading to the independent band invariants $\omega$, $\tilde{\omega}$. In terms of the decompositions (\ref{eq:UF1},\ref{eq:UF2}) and the form of $F$ in the two-band case (\ref{eq:Fcomponents}), we find that
\begin{align}
    &q=2\eta\zeta, &\tilde{q}=2\eta\zeta^*,
\end{align}
which gives
\begin{align}
    \omega,\tilde{\omega}=\nu_0\pm\nu_\pi.\label{eq:omeganunu}
\end{align}
The difference in the sign multiplying $\nu_\pi$ guarantees that these equations are preserved under the half-period shift since $\tilde{\nu}_\pi=-\nu_\pi$ \eqref{eq:halfperiodnu}. We also note that one can use the inverse relations of (\ref{eq:omeganunu}) to formulate the bulk-boundary correspondence in terms of the band invariants,
\begin{align}
    &N_0=\frac{\omega+\tilde{\omega}}{2}, &N_\pi=\frac{\omega-\tilde{\omega}}{2}.
\end{align}
This formulation gives a new interpretation of the band invariants:
\begin{align}
    &\omega,\tilde{\omega}=N,\tilde{N}, &\text{where}\quad N,\tilde{N}=N_0\pm N_\pi\label{eq:N12}
\end{align}
gives the total number of localized edge states at times $0$ and $T/2$.

The gapless case corresponds to vanishing $q$, so that the invariants \eqref{eq:omeganunu} become ill-defined. Again, one can define the analytic continuation $q(z)$ and the regularized invariants
\begin{align}
    &\omega^\pm=\lim_{\delta\to 0}\frac{1}{2\pi i}\oint_{S^1(1\pm\delta)}\frac{dq}{q},
\end{align}
and similarly for $\tilde{\omega}^\pm$. Then we find that the relationship with the gap invariants is modified to
\begin{align}
    &\omega^\pm=\nu_0^\pm+\nu_\pi^\pm,\\
    &\tilde{\omega}^\pm=\nu_0^\pm-\nu_\pi^\mp,
\end{align}
where in particular the contribution of $\nu_\pi$ to $\tilde{\omega}$ is swapped between $\nu_\pi^-$ and $\nu_\pi^+$. This swap is due to the same reason as in the transformation law \eqref{eq:halfperiodnu} above. Here, it implies that the interpretation of the regularized band invariants as the instantaneous total number of edge states per boundary, equation \eqref{eq:N12}, is also valid in the gapless case.

In Appendix \ref{app:geometryinvariants}, we connect the identities \eqref{eq:omeganunu} with the geometric picture of the invariants as linking numbers between oriented loops in the sphere, where the key is a correspondence between the chiral decomposition (\ref{eq:UF1},\ref{eq:UF2}) and the geometry of the Hopf map $S^3\to S^2$.

\section{Examples}
 \label{sec:examples}

Consider the Hamiltonians
\begin{align}
    H_\alpha = \begin{pmatrix}
        0 & T^{\alpha}\\
        T^{-\alpha} &0
    \end{pmatrix},\label{eq:HalphaTalpha}
\end{align}
which correspon to hopping between sites of opposite chirality which lie $\alpha$ unit cells away from each other (Fig. \ref{fig:chirallattice}). Since $H_\alpha$ anticommutes with $\Gamma$, one can construct Floquet evolutions which satisfy the dynamical chiral symmetry \eqref{eq:chiralsym0} by multiplying the $H_\alpha$ by an even function of time, as we do below. In the bulk, we have
\begin{equation}
    H_\alpha = \sum_k\begin{pmatrix}
        0 & e^{-i\alpha k}\\
        e^{i\alpha k} & 0
    \end{pmatrix}|k\rangle\langle k|,\label{eq:halphakk}
\end{equation}
which gives the gapped static spectrum $E_\alpha=\pm 1$ with flat bands. With open boundary conditions, one can check directly from \eqref{eq:HalphaTalpha} that there are $N=\alpha$ zero modes per boundary: since $H_\alpha$ only involves hoppings to the sites of opposite chirality that are $\alpha$ unit cells away, the first $|\alpha|$ sites of the $A$ sublattice are free on the left edge, as well as the last $|\alpha|$ sites of the $B$ sublattice on the right edge (Fig. \ref{fig:chirallattice}) if $\alpha>0$, and oppositely if $\alpha<0$. In particular, $t_0H_0+t_1H_1$ corresponds to the Su-Schrieffer-Heeger (SSH) chain, with the topological regime being $t_1>t_0$. Thus the linear combinations of the $H_\alpha$ are generalizaions of the SSH model to the class of chiral chains. In the static limit, these can display zero modes even in the gapless regime \cite{verresen2018topology}. As we now discuss, one can drive these Hamiltonians into the gapless Floquet phases we described above.

\begin{figure}[]
\centering
\includegraphics[width=\linewidth]{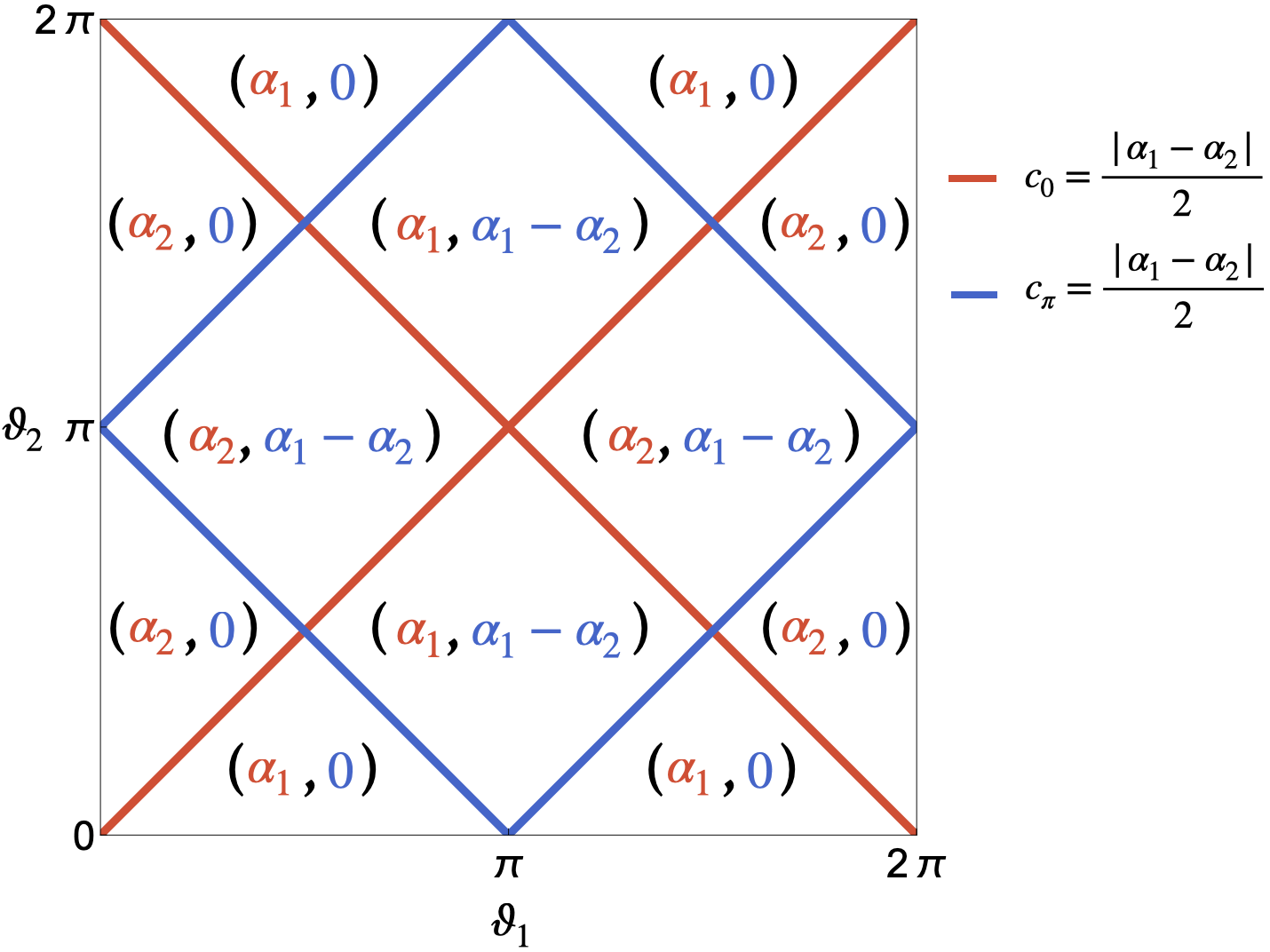}
\caption{Phase diagram for the two-step Floquet drive \eqref{eq:Fbinary} with $\alpha_1>\alpha_2>0$. The bulk spectrum is gapped, except on the red lines, where the zero-quasienergy gap closes, and on the blue lines, where the $\pi$-quasienergy gap closes. In the gapped regions, $(N_0,N_\pi)$ denotes the number of zero- and $\pi$-modes per boundary. On the red lines, there are $\alpha_2$ zero-modes, even though the bulk is gapless. On the blue lines, there are no $\pi$ modes.}
\label{fig:phase12F}
\end{figure}

\subsection{Zero-modes in the gapless case}

Consider a two-step drive, where half of the period $T$ is governed by the time-independent Hamiltonian $H_{\alpha_1}$ and the other half by $H_{\alpha_2}$. We choose the time origin $t=0$ so that the Hamiltonian is even in time,
\begin{equation}
    H(t) = \begin{cases}
        t_1 H_{\alpha_1}, & 0<t({\rm mod}\,T)<\frac{T}{4},\\
        t_2 H_{\alpha_2}, & \frac{T}{4}<t({\rm mod}\,T)<\frac{3T}{4},\\
        t_1 H_{\alpha_1}, & \frac{3T}{4}<t({\rm mod}\,T)<T,\label{eq:binaryHt}
    \end{cases}
\end{equation}
where the steps have different hopping ranges $\alpha_{1,2}$ and different hopping strengths $t_{1,2}$. Then the half-period evolution is given by
\begin{equation}
    F = e^{-i \frac{\vartheta_2}{2} H_{\alpha_2}}e^{-i \frac{\vartheta_1}{2} H_{\alpha_1}},\label{eq:Fbinary}
\end{equation}
where $\vartheta_{1,2}=t_{1,2}T/2$ are dimensionless hopping parameters. Since the overall sign of $F$ is irrelevant, the spectrum depends on $\vartheta_{1,2}$ periodically with a period $2\pi$. Comparison with the parametrization \eqref{eq:Fcomponents} gives
\begin{align}
    &\zeta(z) = \cos\frac{\vartheta_1}{2}\cos\frac{\vartheta_2}{2}-\sin\frac{\vartheta_1}{2}\sin\frac{\vartheta_2}{2} z^{\alpha_1-\alpha_2},\\
    &\eta(z) = -i\left(\sin\frac{\vartheta_1}{2}\cos\frac{\vartheta_2}{2} z^{\alpha_1}+\sin\frac{\vartheta_2}{2}\cos\frac{\vartheta_1}{2} z^{\alpha_2}\right).
\end{align}

Let us consider the case $\alpha_1>\alpha_2>0$. Then $\eta(z)$ has a zero of order $\alpha_2$ at the origin, plus $\alpha_1-\alpha_2$ zeros given by
\begin{equation}
    z_*^{\alpha_1-\alpha_2} = -\cot\frac{\vartheta_1}{2}\tan\frac{\vartheta_2}{2}.
\end{equation}
The gap at quasienergy $0$ is closed when this expression is plus or minus one, which happens on the lines
\begin{align}
    \vartheta_2 = \pm\vartheta_1.\label{eq:0gaplessbinary}
\end{align}
The resulting phase diagram is shown in Fig.~\ref{fig:phase12F}. Away from the lines \eqref{eq:0gaplessbinary}, the bulk is gapped, and as shown in Fig.~\ref{fig:phase12F}, there are either $N_0=\nu_0^-=\alpha_1$ or $N_0=\nu_0^-=\alpha_2$ zero-modes per boundary. Note that these regions include the limits $\vartheta_2\to 0$ and $\vartheta_1\to 0$, respectively, which are related to the static evolutions given by the Hamiltonians $H_{\alpha_1}$ and $H_{\alpha_2}$. Thus, the zero-modes in each region are continuously connected to the zero-modes in the static case. As expected, the Floquet dynamics can close the bulk gap, but surprisingly some of the zero-modes survive. For example, on the line $\vartheta_2=\vartheta_1$, there are
\begin{equation}
    N_0=\nu_0^-=\alpha_2
\end{equation}
zero-modes, even though there are
\begin{equation}
    2c_0=\nu_0^+-\nu_0^-=\alpha_1-\alpha_2>0
\end{equation}
Dirac points in the bulk spectrum. This model shows that Floquet phases which do not have a gap at zero quasi-energy can also present topologically-protected zero-mode states, whose numbers are given by the doubled invariants $\nu_0^\pm$.

We consider now the $\pi$ gap. For $\alpha_1>\alpha_2>0$, $\zeta(z)$ has $\alpha_1-\alpha_2$ roots given by
\begin{equation}
    z_*^{\alpha_1-\alpha_2} = \cot\frac{\vartheta_1}{2}\cot\frac{\vartheta_2}{2}.
\end{equation}
Thus the gap at quasienergy $\pi$ is closed when this expression is equal to plus or minus one, which corresponds to the lines
\begin{align}
    \vartheta_2 = \pm(\vartheta_1-\pi).\label{eq:pigaplessbinary}
\end{align}
Away from these lines, the system is gapped and has either $N_\pi=\nu_\pi^-=0$ or $N_\pi=\nu_\pi^-=\alpha_1-\alpha_2$, as shown in the phase diagram (Fig. \ref{fig:phase12F}). On the gapless lines \eqref{eq:pigaplessbinary}, one finds that
\begin{equation}
    N_\pi=\nu_\pi^-=0,
\end{equation}
so that there are no $\pi$-modes when the bulk $\pi$ gap is closed. The bulk spectrum has
\begin{equation}
    c_\pi = \frac{\nu_\pi^+-\nu_\pi^-}{2} = \frac{\alpha_1-\alpha_2}{2}>0.\label{eq:cpibinary}
\end{equation}
Thus in this simple model of a binary drive, we only find topologically trivial $\pi$-gapless lines, with no localized $\pi$ modes.

\begin{figure}[]
\centering
\includegraphics[width=0.9\linewidth]{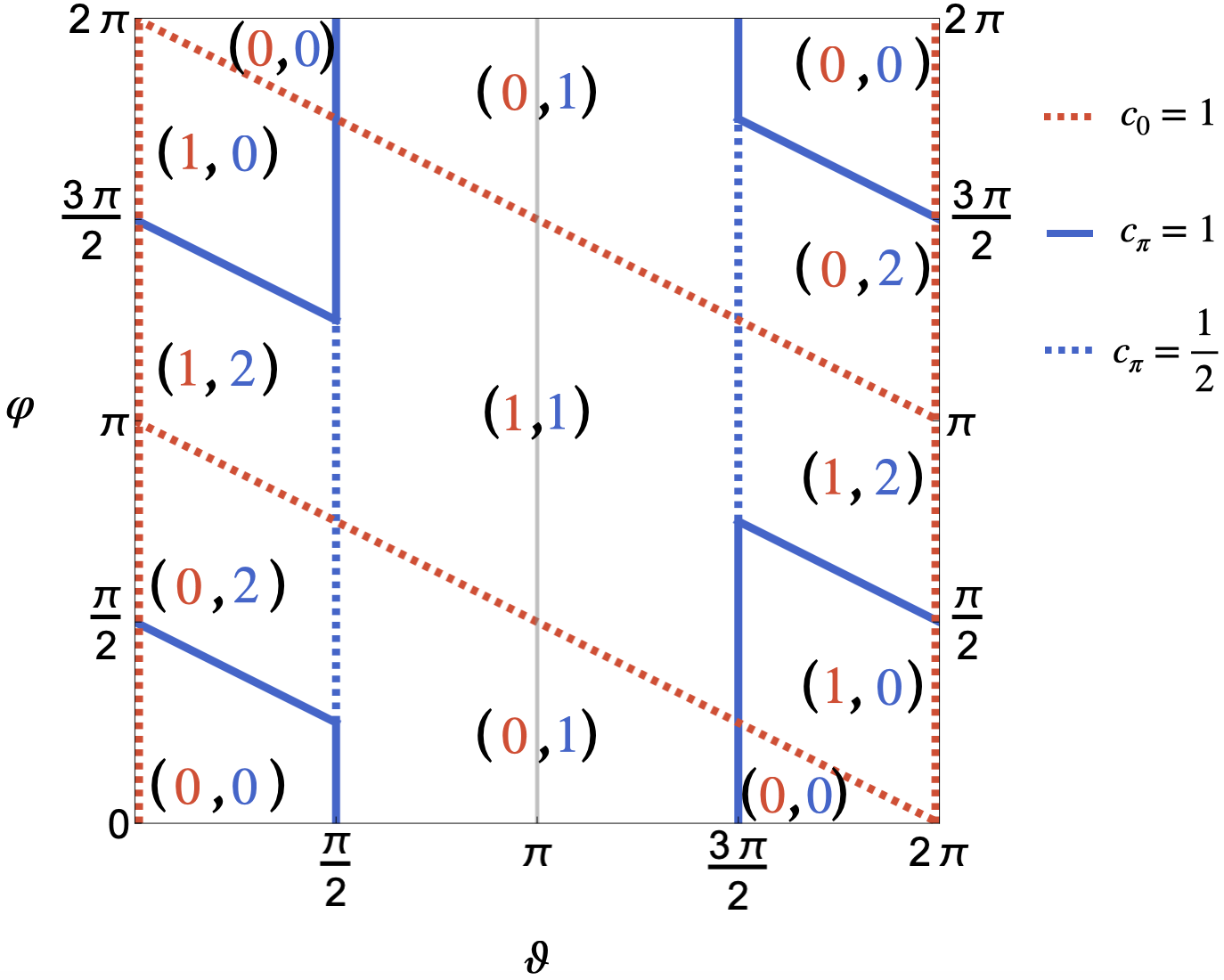}
\caption{Phase diagram for the four-step drive \eqref{eq:fourstepF}. In the whole plane, the bulk gap at zero-quasienergy is closed, $c_0=\frac{1}{2}$. On the red dashed lines, there is an extra Dirac point at zero-quasienergy in the bulk, so that $c_0=1$. The bulk gap at $\pi$-quasienergy is closed on the blue lines, with $c_\pi=1$ on the solid lines and $c_\pi=\frac{1}{2}$ on the dashed lines. The number of zero- and $\pi$-modes $(N_0, N_{\pi})$ on each region is outlined. In particular, along the $\vartheta=\frac{3\pi}{2}$ line, the bulk is gapless at both zero- and $\pi$-quasienergy, but the number of zero- and $\pi$-modes per boundary change, as shown in detail in Fig.~\ref{fig:localization}. Interestingly, the bulk displays flat bands at $\vartheta=\pi$ (gray line).}
\label{fig:four-step-phase}
\end{figure}

\subsection{$\pi$-modes in the gapless case}
\label{sec:pi-gapless}

As an example of localized $\pi$-modes in the gapless case, we consider the slightly more sophisticated Floquet evolution
\begin{equation}
    F=e^{i\frac{\vartheta+\varphi}{2}H_0}e^{i\frac{\vartheta}{2}H_1}e^{-i\frac{\varphi}{2}H_2},\label{eq:fourstepF}
\end{equation}
which corresponds to a four-step evolution with hoppings of up to next-to-nearest neighbors. The parametrization \eqref{eq:Fcomponents} corresponding to \eqref{eq:fourstepF} is
\begin{align}
    &\zeta(z) = \frac{1}{4}\{(1+z)^2\cos\vartheta+(1-z)^2\nonumber\\
    &\hspace{2.5cm}+(1-z^2)[\cos\varphi+\cos(\vartheta+\varphi)]\},\\
    &\eta(z) = \frac{i}{2}(1+z)\cos\frac{\vartheta}{2}\Big[(1+z)\sin\frac{\vartheta}{2}\nonumber\\
    &\hspace{3.5cm}+(1-z)\sin\left(\frac{\vartheta}{2}+\varphi\right)\Big].
\end{align}

\begin{figure}[]
\centering
\includegraphics[width=0.9\linewidth]{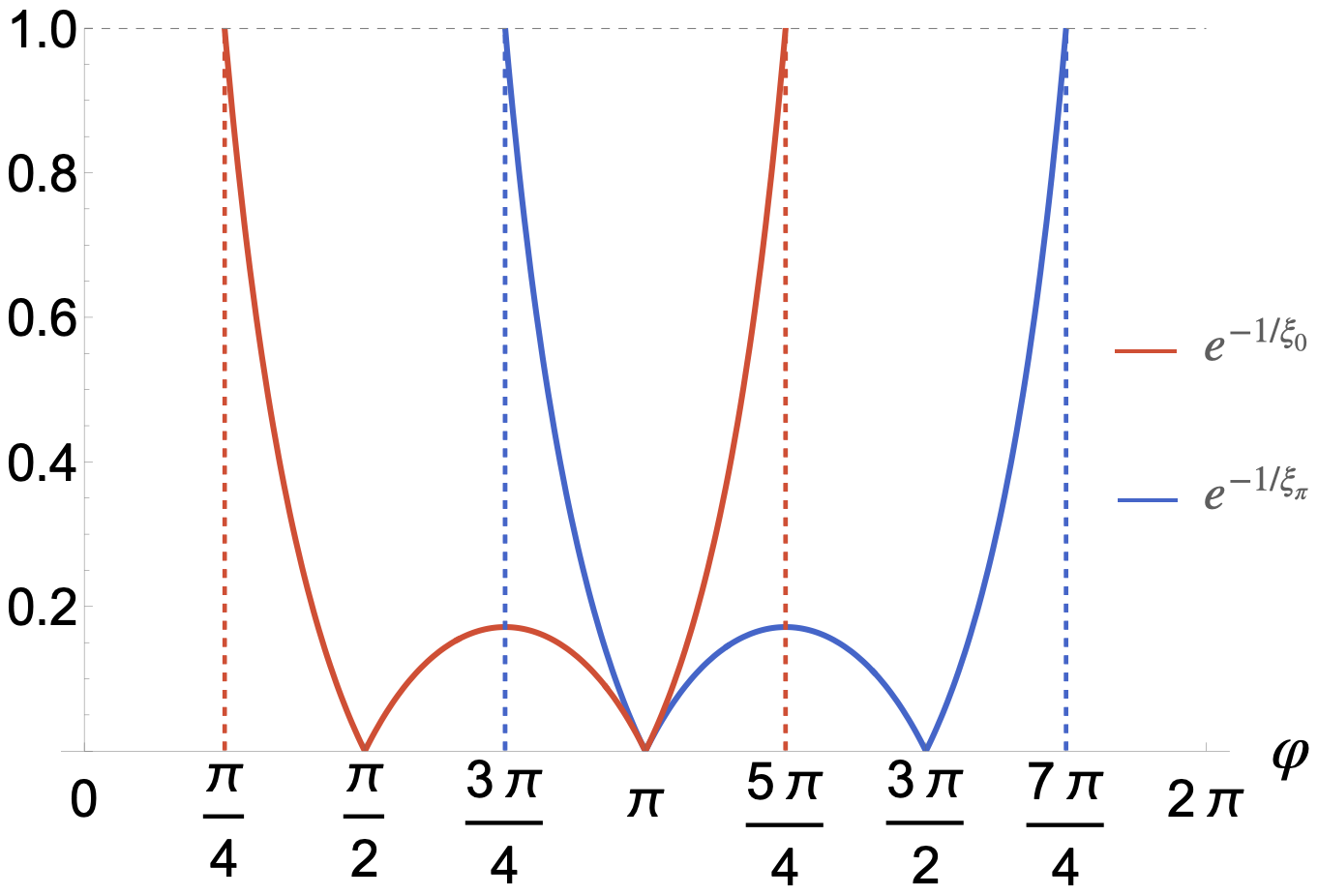}
\caption{Spatial decay $e^{-1/\xi_{0,\pi}}$ of zero- and $\pi$-modes for the four-step drive \eqref{eq:fourstepF} on the line $\vartheta=3\pi/2$. Although there are no bulk gaps at $0$ or $\pi$ quasienergies on the whole line  $\vartheta=3\pi/2$, one finds transitions where the number of zero- and $\pi$-modes $(N_0,N_\pi)$ change as $(0,0)\mapsto (1,0)\mapsto (1,1)\mapsto (0,1)\mapsto (0,0)$. At the transitions, the localization length diverges.}
\label{fig:localization}
\end{figure}

Applying the same analysis as in the previous example, one finds the phase diagram in Fig.~\ref{fig:four-step-phase}. In particular, consider the line
\begin{equation}
    \vartheta = \frac{3\pi}{2}, \varphi\in [0,2\pi),
\end{equation}
on which $(\zeta,\eta)$ simplify to
\begin{align}
    &\zeta(z) = \frac{1-z}{4}[1-z+(1+z)(\cos\varphi+\sin\varphi)],\\
    &\eta(z) = -\frac{i(1+z)}{4}\left[(1+z)+\sqrt{2}(1-z)\cos\left(\varphi+\frac{\pi}{4}\right)\right].
\end{align}
As $\eta(e^{i\pi})=\zeta(e^{i 0})=0$, we see that the bulk spectrum is gapless both at $0$ and $\pi$ quasienergies. However, there are extra roots which lead to edge states. Namely, $\zeta(z)$ has a root
\begin{equation}
    z_\varphi = -1-\frac{2}{\sin\varphi+\cos\varphi-1},\label{eq:zphi}
\end{equation}
with absolute value
\begin{align}
    e^{-1/\xi_\pi}=\Big|1+\frac{2}{\sin\varphi+\cos\varphi-1}\Big|,\label{eq:xipiquartic}
\end{align}
which gives a localized $\pi$-mode for $\phi\in (3\pi/4,7\pi/4)$, and $\eta(z)$ has a root with absolute value
\begin{align}
    e^{-1/\xi_0}=\Big|1-\frac{2}{\sin\varphi-\cos\varphi+1}\Big|,\label{eq:xizeroquartic}
\end{align}
which gives a localized zero-mode for $\phi\in (\pi/4,5\pi/4)$. Thus, the line $\vartheta=3\pi/2$ displays transitions between gapless Floquet phases, where the numbers of zero- and $\pi$-modes change as $(N_0,N_\pi)=(0,0)\mapsto (1,0)\mapsto (1,1)\mapsto (0,1)\mapsto (0,0)$ as $\varphi$ goes from $0$ to $2\pi$ (see Fig.~\ref{fig:localization}). In Figures \eqref{fig:Fkwinding} d) and e), we show the winding diagrams of $F(k)$ in the topologically trivial case $\varphi=0$ (no zero- or $\pi$-modes), and in the topologically non-trivial case $\varphi=\pi$ (one zero- and one $\pi$-mode), respectively.

\begin{figure*}
    \includegraphics[width=0.32\textwidth]{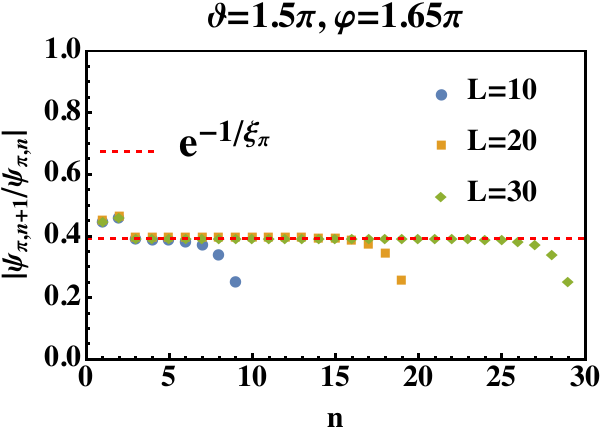}
    \includegraphics[width=0.32\textwidth]{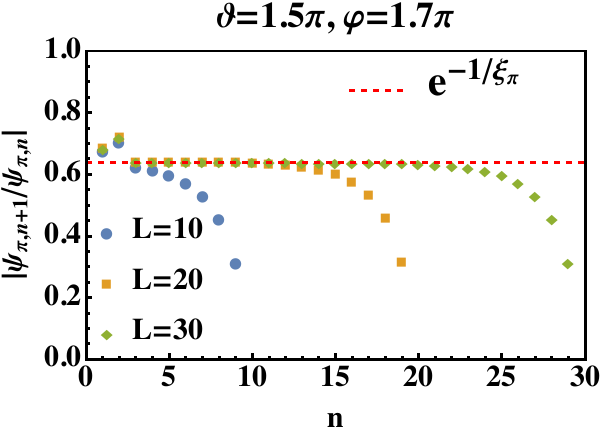}
    \includegraphics[width=0.32\textwidth]{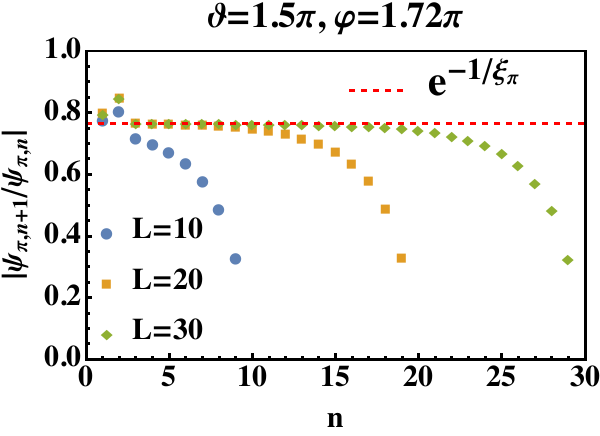}

    \caption{Amplitude ratio of the $\pi$-mode operator on different Majoranas along the chain, obtained from ED in the Majorana basis, and for a chain of $L$ sites. The amplitude ratio quickly approaches the value $e^{-1/\xi_\pi}$ predicted from the roots of the $\zeta$ component of the $F$ matrix, while the amplitudes of the mode on the first few sites are fixed by boundary conditions.}
    \label{Fig: localization length}
\end{figure*}

\subsection{Fermionic SPT phases}\label{sec:fermions}

The chiral Floquet evolutions we discuss are related to the AIII and BDI classes of Floquet symmetry-protected topological phases in one dimension.

\paragraph{AIII class.} In the AIII class, one considers Hamiltonians
\begin{equation}
    \hat{H}(t) = \sum_{n,m}\hc_n^\dagger H_{n,m}(t)\hc_m,
\end{equation}
where the single-particle Hamiltonian $H$ satisfies the chiral symmetry \eqref{eq:chiralsym0}. Thus, one can construct the invariants and topological boundary states which we discussed here. In particular, given the wavefunctions for topological edge states of the single-particle Hamiltonian $|\psi_{0/\pi}\rangle = \sum_n\psi_{0/\pi,n}|n\rangle$, one can construct the zero- and $\pi$-mode operators
\begin{align}
    \hat{\Psi}_{0/\pi}^\dagger = \sum_n\psi_{0/\pi,n}\hc_n^\dagger,
\end{align}
which satisfy the time-evolution
\begin{align}
    &[\hat{U}_\tau,\hat{\Psi}_0^\dagger]=0, &\{\hat{U}_\tau,\hat{\Psi}_\pi^\dagger\}=0.\label{eq:conservedpsi}
\end{align}

\paragraph{BDI class.} Similarly, in the BDI class one considers Hamiltonians
\begin{equation}
    \hat{H} = \frac{1}{4}\sum_{n,m}\ha_n^\dagger H_{n,m}\ha_m,
\end{equation}
where the $\ha_n$ are Majorana operators and the single-particle Hamiltonian $H$ satisfies the chiral symmetry \eqref{eq:chiralsym0}. The zero- and $\pi$-mode operators are given by
\begin{align}
    \hat{\Psi}_{0/\pi} = \sum_n\psi_{0/\pi,n}\ha_n,
\end{align}
which satisfy equations (\ref{eq:conservedpsi}). In Appendix \ref{app:freefermions}, we give more details of this correspondence relevant to the driven case.

In particular, both the driven SSH chain,
\begin{equation}
    H(t) = \begin{cases}
        -\mu \sum_n\hc_{A,n}^\dagger\hc_{B,n}+\text{h.c.}, & 0<t<\frac{T}{2},\\
        2t \sum_n\hc_{A,n+1}^\dagger\hc_{B,n}+\text{h.c.}, & \frac{T}{2}<t<T,
    \end{cases},
\end{equation}
and the driven Kitaev chain,
\begin{equation}
    H(t) = \begin{cases}
        i\frac{\mu}{2}\sum_n\ha_{A,n}\ha_{B,n}, & 0<t<\frac{T}{2},\\
        -it \sum_n\ha_{A,n+1}\ha_{B,n}, & \frac{T}{2}<t<T,
    \end{cases}\label{eq:kitaev},
\end{equation}
correspond to the two-step drive \eqref{eq:binaryHt} by a shift $t\mapsto t+T/4$ and, therefore, display zero- and $\pi$-modes according to the phase diagram \ref{fig:phase12F}, with $(\alpha_1,\alpha_2)=(0,1)$, $\vartheta_1 = -\mu T/2$, and $\vartheta_2=t T$. Let us now construct related models displaying a localized edge $\pi$-mode in the gapless regime for spin-chain Hamiltonians.

\subsection{Spin chains and the effect of interactions}

It is interesting to explore the many-body aspects of the zero- and $\pi$-modes in the spin language. For the Majorana chain, one can use the Jordan-Wigner transformation
\begin{align}
    &\ha_{A,n} = \left(\prod_{j=1}^{n-1}\sigma^z_j\right)\sigma^x_n, &\ha_{B,n} = \left(\prod_{j=1}^{n-1}\sigma^z_j\right)\sigma^y_n.
\end{align}
This transformation maps the driven Kitaev chain \eqref{eq:kitaev} to the driven transverse field Ising model,
\begin{equation}
    \hat{F} = e^{-i \frac{\vartheta_2}{4}\sum_n\sigma^z_n}e^{i \frac{\vartheta_1}{4} \sum_n\sigma^x_n\sigma^x_{n+1}},\label{eq:Fbinaryspin}
\end{equation}
for which the stability of zero- and $\pi$-modes against interactions was studied in \cite{Yates19,yeh2023decay}. Likewise, by taking $\alpha_1=1$, $\alpha_2=-1$, we obtain a model related to the $XY$ chain,
\begin{equation}
    \hat{F} = e^{i \frac{\vartheta_1}{4} \sum_n\sigma^y_n\sigma^y_{n+1}}e^{i \frac{\vartheta_1}{4} \sum_n\sigma^x_n\sigma^x_{n+1}},\label{eq:Fbinaryspin2}
\end{equation}
which, therefore, also has a phase diagram given by Fig.~\ref{fig:phase12F}.

To study the $\pi$-mode in the gapless regime, we can consider the spin version of \eqref{eq:fourstepF} which becomes
\begin{align}
    \hat{F}=&e^{i\frac{\vartheta+\varphi}{4}\sum_n\sigma^z_n} e^{-i\frac{\vartheta}{4}\sum_n\sigma^x_n\sigma^x_{n+1}}e^{i\frac{\varphi}{4}\sum_n\sigma^x_n\sigma^z_{n+1}\sigma^x_{n+2}}.\label{eq:Ffoursigma}
\end{align}
As we discussed, the bulk system is gapless both at $0$ and $\pi$ quasienergies for $\vartheta=3\pi/2$ but, for $\varphi\in (3\pi/4,7\pi/4)$, it has an edge $\pi$-mode. It is asymptotically (at large $n$) given by
\begin{align}
    \hat{\Psi}_\pi &\sim \mathcal{N}_\varphi\sum_{n=1}^\infty z_\varphi^{n-1}\ha_{A,n}\label{eq:psipiahat}\\
    &=\mathcal{N}_\varphi\sum_{n=1}^\infty z_\varphi^{n-1}\left(\prod_{j=1}^{n-1}\sigma^z_j\right)\sigma^x_n,\label{eq:psipisigma}
\end{align}
where $z_\varphi$ is given by equation \eqref{eq:zphi} and $\mathcal{N}_\varphi$ is a normalization constant. By exact diagonalization (ED) of a finite chain, in the Majorana basis of the Floquet unitary \eqref{eq:Ffoursigma}, we find that there is a $\pi$-mode localized on the left. The ratio between its amplitudes on subsequent sites is given by $e^{-1/\xi_\pi}$ everywhere except on the first two sites, which are subject to boundary conditions, see Fig.~\ref{Fig: localization length}.

\begin{figure}[]
\centering
    \includegraphics[width=0.4\textwidth]{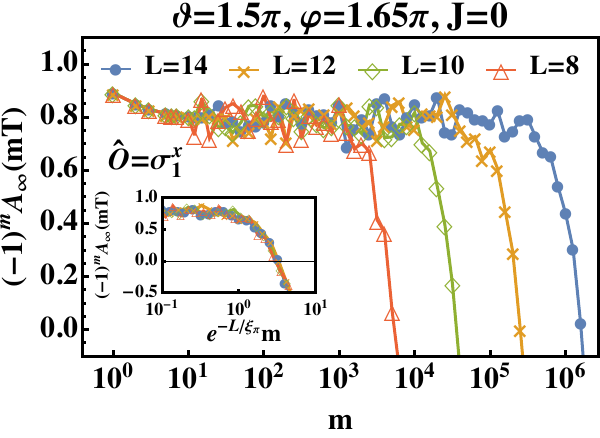}
    \includegraphics[width=0.4\textwidth]{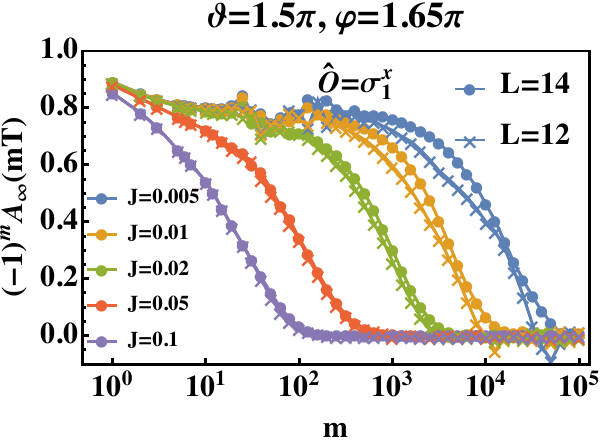}
    \includegraphics[width=0.4\textwidth]{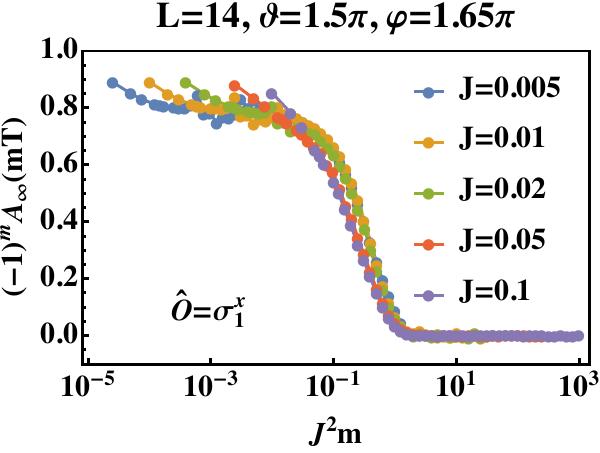}

    \caption{The ED results for the autocorrelation of $\sigma_1^x$ under time evolution \eqref{eq:perturbation} with $\vartheta = 1.5\pi$ and $\varphi=1.65\pi$, where only one $\pi$-mode exists in the non-interacting limit $J=0$. Top panel: System size effect in the non-interacting limit. The inset shows that the results for different system sizes $L = 8, 10, 12 ,14$ collapse via time rescaling $e^{-L/\xi_\pi}m$. Hence, the lifetime of the $\pi$-mode is exponential in system size and it is an exact conserved operator in the thermodynamic limit. Middle panel: Autocorrelation with perturbation  $J$ for system size $L = 12,14$. In the presence of interactions, the lifetime of the $\pi$-mode becomes $L$-independent for large system sizes. 
    The numerical results only contain $L = 12,14$ since for shorter lengths the non-interacting lifetime is shorter than the interaction induced lifetime. Bottom panel: The rescaled autocorrelation for $L=14$ from the middle panel. The data collapses for small values of the perturbation $J$ and indicates a decay rate that is second order in $J$.}
    \label{Fig: autocorrelation} 
\end{figure}

From equation \eqref{eq:conservedpsi}, we know that in the thermodynamic limit the $\pi$-mode is a conserved operator. Due to the localization of \eqref{eq:psipisigma} on the left boundary, its presence can be probed even in a finite chain using the dynamical autocorrelation function
\begin{equation}
    A_\infty(mT) = \frac{1}{2^L}\Tr[\hat{O}(mT)\hat{O}(0)],\label{eq:ainftya}
\end{equation}
which has a long-lived plateau depending on the overlap between the edge modes and the Hermitian operator $\hat{O}$. Usually $\hat{O}$ is chosen to be $\sigma_1^x$ (the first Majorana) as it is expected to have the largest overlap with edge modes, however see Appendix \ref{app:Numerics} for an exception. In the top panel of Fig.~\ref{Fig: autocorrelation}, we consider $\vartheta = 1.5\pi$ and $\varphi = 1.65\pi$, where only one $\pi$-mode exists as shown in Fig.~\ref{fig:localization}. Numerically, it is convenient to consider $(-1)^m A_\infty(mT)$ so that the period-doubled oscillations of the $\pi$-mode are smoothened out. The inset plot in the top panel of Fig.~\ref{Fig: autocorrelation} illustrates that the lifetime of the $\pi$-mode collapses on rescaling time $m$ by $e^{-L/\xi_\pi}m$. Namely, the lifetime of the plateau grows exponentially with the system size $L$, which agrees with the expectation that the decay comes from tunneling processes that hybridize the edge modes at the two ends of the finite chain. Note that this is true even in the absence of a gap in the quasienergy spectrum. 

We also consider the effect of interactions, such that the time-evolution is given by
\begin{align}
    \hat{U} =& e^{i\frac{\varphi}{4}\sum_n\sigma^x_n\sigma^z_{n+1}\sigma^x_{n+2}} e^{iJ\sum_n\sigma^z_n\sigma^z_{n+1}}e^{-i\frac{\vartheta}{4}\sum_n\sigma^x_n\sigma^x_{n+1}} \nonumber\\
    &\times e^{i\frac{\varphi+\vartheta}{2}\sum_n\sigma^z_n} e^{-i\frac{\vartheta}{4}\sum_n\sigma^x_n\sigma^x_{n+1}}e^{iJ\sum_n\sigma^z_n\sigma^z_{n+1}}\nonumber\\
    &\times e^{i\frac{\varphi}{4}\sum_n\sigma^x_n\sigma^z_{n+1}\sigma^x_{n+2}}.
    \label{eq:perturbation}
\end{align}
Above, the $\sigma_n^z\sigma_{n+1}^z$ term corresponds to a four-Majorana interaction. Without the perturbation, $J = 0$, the half-period time evolution is reduced to  \eqref{eq:Ffoursigma}. The middle panel in Fig.~\ref{Fig: autocorrelation} shows that the autocorrelation function decays faster 
as the strength of the perturbation $J$ increases. Since the quasienergy spectrum is gapless, one expects a decay rate that obeys Fermi's Golden Rule in being second order in the strength of the perturbation, and this is supported by the bottom panel in Fig.~\ref{Fig: autocorrelation}.
One could obtain decay rates that are higher order in the perturbation, but this occurs typically for flat quasienergy bands or special circumstances leading to vanishing matrix elements of the perturbation \cite{yeh2023slowly,yeh2023decay}.

When both gapless zero- and $\pi$-modes exist, the two edge modes could be maximally peaked at different sites of the Majorana chain. In Appendix \ref{app:Numerics}, we consider the case with $\vartheta = 1.5\pi$ and $\varphi = 0.85\pi$, where the $\pi$-mode has the same localization length as $\vartheta = 1.5\pi$ and $\varphi = 1.65\pi$ studied above, but the amplitude is maximally peaked at the fifth Majorana. In contrast, the  zero-mode has maximal amplitude at the first Majorana. We observe second order decay rates for both the zero- and $\pi$-modes under the perturbed unitary \eqref{eq:perturbation}. Although the localization length is different for zero- and $\pi$-modes in this example, they share similar decay rates. Qualitatively, the decay channel from scattering between edge modes and bulk excitations depends on the localization length of the edge modes \cite{yeh2023decay,YehProduct}. Since the decay rates of the zero- and $\pi$-modes are similar, it suggests that the dominant decay channel corresponds to scattering between the two edge modes.

\section{Conclusions}
\label{sec:conclusions}

We explored the generalization of Floquet topology to the gapless case. The notion of topological equivalence is defined by continuous deformations which do not break chiral symmetry or change the relevant central charges. In the case of systems with chiral symmetry, the two relevant gaps are $e^{i0}$, $e^{i\pi}$, with the corresponding central charges $c_{0,\pi}$ counting the number of crossings at these quasienergies by the bulk quasienergy spectrum. In the bulk, we present the topological invariants distinguishing different classes of gapless evolutions. For open boundary conditions, those classes are characterized by the number of topological zero- and $\pi$-modes. We generalize the bulk-boundary correspondence to the gapless Floquet case connecting the bulk invariants with numbers of edge zero- and $\pi$-modes.

The essential part of our construction is the use of the half-period evolution unitary $F$. In chiral evolutions, it completely determines the Floquet spectrum, including the central charges and the topological invariants. In terms of $F$, we found integral expressions for the integer invariants at the $0$ and $\pi$ quasienergies, which are explicitly independent of the choice of basis, number of bands, and the choice of a logarithmic branch cut (which is implicit in constructions in terms of the Floquet Hamiltonian). This formulation is well-suited for generalization to the gapless case, where one uses twice as many invariants. The difference within the pair of invariants determines the central charge of the bulk spectrum, while the smallest one counts the number of topological edge modes.  We also formulated the relation between band and gap invariants, which is valid for both gapped and gapless spectra. In the two-band case, the invariants have a geometric interpretation in terms of the projection of linked loops under the Hopf map.

Our construction is directly related to the AIII and BDI classes of fermionic Floquet SPTs, and we gave the explicit construction of the fermionic zero- and $\pi$-mode operators. The fermionic models can be mapped to spin chains, and we showed how the presence of topological edge modes can be verified in finite-size simulations in terms of the long-time plateau of dynamical autocorrelation functions of boundary operators. We used this technique to verify the analytically computed phase diagrams for examples displaying zero- and $\pi$-modes in the gapless regime. Finally, an investigation of the effect of interactions on the edge modes (both $0$ and $\pi$) of gapless drives showed that these modes acquire a lifetime which is perturbative in the interaction strength. 

Our work leads to a number of interesting new questions. It seems that the generalization of topology and edge modes should also apply to all the classes of Floquet SPTs for which the gap invariants are at least $\mathbb{Z}$-valued. In connection to the problem of strong and almost-strong zero- and $\pi$-modes, further study is needed to assert the effect of interactions in the many-body realizations of these models. Finally, the gapless topology we discuss here is a generalization of extrinsic gapless topology to the Floquet-driven case \cite{thorngren2021intrinsically}. Whether a notion of intrinsically gapless phases also appears in the Floquet case is left for future exploration.

While completing our manuscript, we became aware of \cite{zhou2024floquet}, which presents work overlapping with ours. Similar to us, the authors of \cite{zhou2024floquet}   realized that Floquet driving a critical phase can enrich its topological content, demonstrating this by studying the model referred here as the binary drive, equation \eqref{eq:Fbinary} in the BDI class. Our contribution  goes further by introducing the concept of gapless Floquet phases, extending Floquet topology to gapless regimes. Specifically, we identify the appropriate topological invariants for gapless systems and establish the bulk-boundary correspondence in this context. Additionally, in Sec.~\ref{sec:pi-gapless} we provide a one-dimensional example where not only zero-modes but also $\pi$-modes emerge, when the bulk is gapless at the respective quasienergies.  Further, we study the stability of the gapless edge modes to interactions.

\section{Acknowledgments}

This work was supported by a postdoctoral fellowship of the Tsung-
Dao Lee Institute (GC), by the US Department of Energy, Office of
Science, Basic Energy Sciences, under Award No.~DE-SC0010821 (HCY, AM) and by the National Science Foundation under Grant NSF DMR-2116767 (LK and AGA). HCY acknowledges support of the NYU IT High Performance Computing resources, services, and staff expertise.

We thank M. Braverman for bringing the Ref.~\cite{hebda2007linking} to our attention.

\appendix
\section{Properties of the topological invariants}\label{app:topinvariants}

We study in more detail the invariants (\ref{eq:nutheta},\ref{eq:nuphi}), in particular their general properties for arbitrary number of bands $N$.

\subsection{Proof of topological invariance}

First, note that
\begin{equation}
    \nu_0 = \frac{1}{i\pi N}\Tr\int [F,\Gamma]^{-1}dF = \int A_0,\label{eq:nu0A0}
\end{equation}
where the one form $A_0=A_{0,\mu}dx^\mu$ corresponds to the gauge potential
\begin{align}
    A_{0,\mu} &=\frac{1}{i\pi N}\sum_{i,j}([F,\Gamma]^{-1})^{ij}\p_\mu F^{ji}.
\end{align}
Here we explicitly represented the $SU(N)$-valued map $x^\mu\mapsto (F(x))^{ij}$ in terms of the matrix components $(i,j)$ and the spacetime components $x^\mu$. In the main text, $(x^\mu) = (k)$ corresponds to the one-dimensional Brillouin zone momentum. We check that $A_0$ is a closed form,
\begin{align}
    dA_0 &= \frac{1}{i\pi N}\Tr(-[F,\Gamma]^{-1}[dF,\Gamma][F,\Gamma]^{-1}dF)\\
    &=\frac{1}{i\pi N}\Big[\Tr([F,\Gamma]^{-1}\Gamma dF[F,\Gamma]^{-1}dF)\label{eq:dA0calculation}\\
    &\hspace{1.5cm}-\Tr([F,\Gamma]^{-1}dF\Gamma [F,\Gamma]^{-1}dF)\Big].\nonumber
\end{align}
Now note that
\begin{align}
    &\Gamma[F,\Gamma]=\Gamma F\Gamma-F = -[F,\Gamma]\Gamma\\
    &\hspace{1cm}\Rightarrow [F,\Gamma]^{-1}\Gamma = -\Gamma[F,\Gamma]^{-1},
\end{align}
where we used that $\Gamma^2=\id$. Substituting in \eqref{eq:dA0calculation}, we can show that
\begin{align}
    &\Tr([F,\Gamma]^{-1}\Gamma dF[F,\Gamma]^{-1}dF)\nonumber\\
    &\hspace{2cm}=-\Tr(\Gamma[F,\Gamma]^{-1} dF[F,\Gamma]^{-1}dF)\\
    &\hspace{2cm}=\Tr([F,\Gamma]^{-1}dF\Gamma[F,\Gamma]^{-1} dF),
\end{align}
where in the second line we used the cyclic property of the trace together with the anti-commutativity of the exterior product of differential forms. Substituting this result in \eqref{eq:dA0calculation}, we find that
\begin{equation}
    dA_0 = 0.
\end{equation}
It follows that the integral \eqref{eq:nu0A0} is a topological invariant. Indeed, suppose that $\gamma_1$ and $\gamma_2$ are two loops in $SU(N)$ which can be continuously deformed into each other. Then there is a surface $\sigma$ whose boundary $\p\sigma$ is given by the loops $\gamma_1$ and $\gamma_2$, where one of the loops has opposite orientation. It follows that
\begin{align}
    \oint_{\gamma_1}A_0 - \oint_{\gamma_2}A_0 = \int_{\p\sigma}A_0 = \int_\sigma dA_0 = 0,
\end{align}
where we used the generalized Stokes' theorem. In other words, slightly perturbed Floquet evolutions, which lead to slightly different loops $F(k)$ in $SU(N)$, give the same value of $\nu_0$. One can similarly define
\begin{align}
    &\nu_\pi = \int A_\pi, &A_{\pi,\mu} = \frac{1}{i\pi N}\sum_{i,j}(\{F,\Gamma\}^{-1})^{ij}\p_\mu F^{ji},
\end{align}
and, using the identity $\{F,\Gamma\}^{-1}\Gamma = \Gamma\{F,\Gamma\}^{-1}$, prove that
\begin{equation}
    dA_\pi = 0.
\end{equation}

\subsection{Winding numbers for general $N$}

The gauge potentials $A_{0,\pi}$ have a nice geometric interpretation which clarifies additional properties. Recall that, from the decompositions (\ref{eq:UF1},\ref{eq:UF2}), the $U(1)$ action $F\mapsto e^{i\chi}F$ acts trivially on the Floquet unitaries $U$, $\tilde{U}$, so that one can consider $F\in SU(N)$. However, the decomposition also reveals another two actions of $U(1)$,
\begin{align}
    &U(1)_0: F\mapsto e^{-i\frac{\theta}{2}\Gamma}F e^{i\frac{\theta}{2}\Gamma},\label{eq:U1theta}\\
    &U(1)_\pi: F\mapsto e^{i\frac{\phi}{2}\Gamma}F e^{i\frac{\phi}{2}\Gamma},\label{eq:U1phi}
\end{align}
which act nontrivially on the Floquet unitaries,
\begin{align}
    U(1)_0:\hspace{.5cm} &U\mapsto e^{-i\frac{\theta}{2}\Gamma}Ue^{i\frac{\theta}{2}\Gamma},\\
    &\tilde{U}\mapsto e^{-i\frac{\theta}{2}\Gamma}\tilde{U}e^{i\frac{\theta}{2}\Gamma},\\
    U(1)_\pi:\hspace{.5cm} &U\mapsto e^{-i\frac{\phi}{2}\Gamma}Ue^{i\frac{\phi}{2}\Gamma}, \\
    &\tilde{U}\mapsto e^{i\frac{\phi}{2}\Gamma}\tilde{U}e^{-i\frac{\phi}{2}\Gamma}.
\end{align}
Note that both give unitary actions on $U$, $\tilde{U}$, which therefore preserve the Floquet spectrum. However, we see that they act differently on the wavefunctions of Floquet eigenstates. The orbits of the $U(1)_0\times U(1)_\pi$ action are isospectral tori, except at the points where the action becomes degenerate.

Consider a path $\theta(s)$ in $SU(N)$ generated by the $U(1)_{0}$ action, which passes through $F(0)$ at $\theta(0)=0$. It is given, to first order, by
\begin{align}
    &F(0)+s\frac{dF}{ds}\Big|_{s=0} = F(0)+s\frac{d\theta}{ds}\Big|_{s=0}\frac{i}{2}[F,\Gamma]\nonumber\\
    &\Rightarrow \frac{d\theta}{ds} = \frac{2}{iN}\Tr\left([F,\Gamma]^{-1}\frac{dF}{ds}\right),
\end{align}
so that the total winding number of the path $\theta(s)$ is given by
\begin{equation}
    \frac{1}{2\pi}\int d\theta = \frac{1}{i\pi N}\Tr\int [F,\Gamma]^{-1}dF = \nu_0.
\end{equation}
Likewise, considering a path $\phi(s)$ generated by the $U(1)_{\pi}$ action, one finds that the winding number of this path is given by
\begin{equation}
    \frac{1}{2\pi}\int d\phi = \frac{1}{i\pi N}\Tr\int \{F,\Gamma\}^{-1}dF = \nu_\pi.
\end{equation}
The fact that these expressions capture the winding in the $(\theta,\phi)$ directions follow from the closeness of these forms (previous subsection). As discussed in the main text, these forms become ill-defined on regions of complex codimension one, which here correspond to the regions where the $U(1)$ actions above become degenerate.

\section{Details of the edge-state construction}\label{app:bulkboundary}

From a given root $z_j$ of $\zeta$, we constructed the positive-chirality state
\begin{equation}
    |\Psi_j\rangle = \begin{pmatrix}
        |\psi_j\rangle\\
        0
    \end{pmatrix},
\end{equation}
where
\begin{equation}
    |\psi_j\rangle = \sum_{n=1}^\infty z_j^{n-1}|n\rangle.
\end{equation}
In the absence of poles of $\zeta(z)$ inside the unit circle, the series $\zeta(z)$ (\ref{eq:zetaetaz}) starts from $z^0$ and we have
\begin{eqnarray}
    &\mathcal{Z}|\psi_j\rangle =\mathcal{Z}&\sum_{n=1}^{\infty} z_j^{n-1}|n\rangle = \sum_{n=1}^{\infty} z_j^{n-1}\sum_{m=0}^\infty \zeta_m|n-m\rangle 
 \\
    &=& \sum_{m=0}^\infty \zeta_m \sum_{n=1}^\infty z_j^{n+m-1}|n\rangle
    =\zeta(z_j)\sum_{n=1}^\infty z_j^{n-1}|n\rangle
    =0, \nonumber
\end{eqnarray}
since $\zeta(z_j)=0$. It follows that
\begin{align}
    F|\Psi_j\rangle =\begin{pmatrix}
        \mathcal{Z}|\psi_j\rangle\\
        \mathcal{N}|\psi_j\rangle 
    \end{pmatrix} =\begin{pmatrix}
        0\\
        \mathcal{N}|\psi_j\rangle 
    \end{pmatrix},
\end{align}
which has negative chirality. After a full period,
\begin{align}
    U|\Psi_j\rangle &= \Gamma F^\dagger\Gamma \begin{pmatrix}
        0\\
        \mathcal{N}|\psi_j\rangle 
    \end{pmatrix}=\begin{pmatrix}
        -\mathcal{N}^\dagger \mathcal{N}\, |\psi_j\rangle  \\
        \mathcal{Z}\mathcal{N}\,|\psi_j\rangle 
    \end{pmatrix}\\
    &=\begin{pmatrix}
        -|\psi_j\rangle\\
        0
    \end{pmatrix}=e^{i\pi}|\Psi_j\rangle,
\end{align}
where we used equations \eqref{eq:ZHcommute}, (\ref{eq:ZZHH}) and (\ref{eq:Zpsizero}). The generalization to the case of repeated roots and when the series expansion of $\zeta(z)$ includes negative powers proceeds as in the static case discussed in \cite{verresen2018topology}.

\section{Geometry of gap and band invariants: the Hopf map}\label{app:geometryinvariants}

Let us dwell here on the geometric meaning of band and gap invariants for the two-band drive $N=2$. 

We start with a general description of the topology of chirally symmetric unitary matrices $U,\tilde U$.
For $N=2$, one can think of the $F$-matrix as an element of the $SU(2)$ group, which is topologically a three-dimensional sphere $S^3$. We parametrize the matrix $F$ as \eqref{eq:Fcomponents} or, even more explicitly, using Hopf coordinates $(\theta,\phi_\pi,\phi_0)$  on $S^3$: 
\begin{align}
    &\zeta=\cos\left(\frac{\theta}{2}\right)e^{i\phi_\pi}, &\eta=\sin\left(\frac{\theta}{2}\right)e^{i\phi_0}.
\end{align}
For each fixed $\theta\in (0,\pi)$, the phases $\phi_{0,\pi}\in [0,2\pi)$ parametrize a torus $\mathbb{T}_\theta$. This torus degenerates into circles $K_{0,\pi}$ for $\theta={0,\pi}$, shown  in Fig.~\ref{fig:Fkwinding}. The gap at the quasienergy $0,\pi$ is closed on those circles, that is, for $\theta=0,\pi$, respectively.

It is easy to compute unitaries $U,\tilde{U}$ using (\ref{eq:UF1},\ref{eq:UF2})  in terms of Hopf parameters. We obtain 
\begin{align}
    &U, \tilde{U}=\cos\theta\id+\sin\theta\sigma(\phi_\pm)\sigma_z, &\phi_\pm=\phi_0\pm\phi_\pi,
 \label{eq:U12polar}
\end{align}
where $\sigma(\phi)=\cos\phi\sigma_x+\sin\phi\sigma_y$. One can think of the angles $(\theta,\phi_\pm)$ as polar coordinates on $S^2$ so that each chirally symmetric unitary $U,\tilde U$, respectively, corresponds to a point on the corresponding $S^2$
\begin{align}
    (\sin\theta\cos\phi_\pm,\sin\theta\sin\phi_\pm,\cos\theta).
 \label{eq:bloch}
\end{align}
Therefore, the space of chiral unitaries is topologically equivalent to $S^2$ and the construction $F\to U,\tilde U$ defines the mapping 
\begin{equation}
\begin{array}{ll}
    U:S^3\to S^2 & \tilde{U}:S^3\to S^2 \\
    \quad\quad F\mapsto \Gamma F^\dagger\Gamma F & \quad\quad F\mapsto F\Gamma F^\dagger\Gamma
\end{array}
\end{equation}
corresponding to the Hopf map and its dual.

The spectrum of both $U$ and $\tilde U$ is given by $e^{\pm i \theta}$. It is the same for $U$ and $\tilde U$ as expected due to \eqref{eq:halfperiodF}. Notice, however, that the corresponding eigenvectors depend on the different phases $\phi_\pm$ and are different for $U$ and $\tilde U$. The requirement of the nonvanishing gap at quasienergies $0$ and $\pi$ translates into the constraint $\theta\neq 0,\pi$ so that eigenvalues of $U,\tilde U$ are different from $\pm 1$. The torus $\mathbb{T}_\theta$ parametrizing the $F$ matrices at a fixed value of $\theta\neq 0,\pi$ is mapped onto a circle at the given latitude $\theta$ of $S^2$ as defined by  \eqref{eq:bloch}.   

So far, the discussion was general, focusing on the topology of chiral unitary matrices. For a translationally invariant one-dimensional system we have as an additional ingredient, the dependence on the quasi-momentum $k\in S^1$. 
This dependence defines two loops on $S^2$ corresponding to $U(k)$ and $\tilde{U}(k)$. The gap condition $\theta \neq 0,\pi$ means that these loops do not cross the north or south poles. The sphere with two removed points has nontrivial fundamental group $\Pi_1=\mathbb{Z}$ and the loops then have well-defined winding numbers corresponding to windings of phases $\phi_\pm(k)$ \eqref{eq:U12polar}. These winding numbers are the band invariants $\omega$, $\tilde{\omega}$. On the other hand, in $S^3$ we have a loop $F(k)$ which is linked with the circles $K_{0,\pi}$ with the linking numbers $\nu_{0,\pi}$ defining gap invariants. We see that the relation between gap and band invariants $\omega,\tilde{\omega}=\nu_0\pm \nu_\pi$ is determined geometrically by the Hopf map of $S^3\to S^2$ conditioned on mapping the two fixed loops $K_{0,\pi}$ in $S^3$ onto north and south poles of $S^2$.

\section{Free fermion models and spin chains}\label{app:freefermions}

The topology we discussed can be straightforwardly mapped to second-quantized free fermion models and spin chains. In particular, the chiral symmetry we discuss corresponds to the AIII and BDI classes of Floquet SPTs.

\subsection{AIII class}
In the AIII class, one considers complex fermion operators $\hc_n$ satisfying the canonical anticommutation relations
\begin{equation}
    \{\hc_n,\hc^\dagger_m\}=\delta_{nm}.
\end{equation}
Chiral symmetry is given by an anti-unitary operator $\hat{S}$ such that
\begin{align}
    &\hat{S} \hc_n \hat{S}^{-1} = \sum_m\Gamma_{n,m}\hc_m^\dagger, &\Gamma^2 = \Gamma^\dagger\Gamma = \id_N.\label{eq:SdefGamma}
\end{align}
In the time-dependent case, the requirement of chiral symmetry is that
\begin{equation}
    \hat{S}\hat{H}(t)\hat{S}^{-1} = \Hat{H}(-t).\label{eq:SchiralH}
\end{equation}
Here, we used the notation $\hat{H}$ to distinguish from the single-particle Hamiltonian $H$, where
\begin{equation}
    \hat{H} = \sum_{n,m}\hc_n^\dagger H_{n,m}\hc_m.
\end{equation}
Using \eqref{eq:SdefGamma} we see that, in terms of the single-particle Hamiltonian, chiral symmetry corresponds to equation \eqref{eq:chiralsym0},
\begin{equation}
    \Gamma H(t)\Gamma = -H(-t),\label{eq:appchiralH}
\end{equation}
with $\Tr H=0$. We also used the anti-linear property of $\hat{S}$ and the Hermiticity of $H$.

The presence of zero- and $\pi$-modes was stated in terms of the single-particle states
\begin{align}
    |\psi_{0/\pi}\rangle = \sum_n\psi_{0/\pi,n}|n\rangle,\label{eq:single-part}
\end{align}
from which we construct the zero- and $\pi$-mode operators,
\begin{align}
    \hat{\Psi}_{0/\pi}^\dagger = \sum_n\psi_{0/\pi,n}\hc_n^\dagger.
\end{align}
We now discuss their main properties. By using the time-dependent Schr\"odinger equation for the single-particle states, one can show that
\begin{align}
    &[\hat{U}_\tau,\hat{\Psi}_0^\dagger]=0, &\{\hat{U}_\tau,\hat{\Psi}_\pi^\dagger\}=0,\label{eq:psi0piU}
\end{align}
where $\hat{U}_\tau$ denotes the Floquet unitary starting from $\tau$ in the free-fermion system. Thus a dynamical signature of the zero- and $\pi$-modes is given by the infinite-temperature boundary correlation function
\begin{equation}
    A_\infty(t) = \frac{1}{2^L}\Tr[\hat{\Psi}_{0,\pi}(\tau+t)\hat{\Psi}_{0,\pi}^\dagger(\tau)].\label{eq:ainftyc}
\end{equation}
At stroboscopic times, it is given by
\begin{align}
    A_\infty(mT) &= (\pm 1)^m\frac{1}{2^{L+1}}\Tr[\{\hat{\Psi}_{0,\pi},\hat{\Psi}_{0,\pi}^\dagger\}]\\
    &=\frac{(\pm 1)^m}{2},
\end{align}
so that the correlation function has a long-lived plateau at stroboscopic times, with alternating sign in the case of a $\pi$-mode. Since the wavefunction \eqref{eq:single-part} is localized close to the boundary, this plateaux can be probed by taking the correlation functions of boundary operators, even if the exact form of the edge mode operator is not known.

The single-particle states have a chirality eigenvalue,
\begin{equation}
    \Gamma|\psi_{0/\pi}\rangle = (-1)^\sigma|\psi_{0/\pi}\rangle,
\end{equation}
for some $\sigma\in\{0,1\}$. For the corresponding operators, it is useful to define a unitary chirality operator
\begin{equation}
    \hat{\Gamma}=\hat{\mathcal{K}}\hat{S},\label{eq:unitarySG}
\end{equation}
where $\hat{\mathcal{K}}$ denotes the complex conjugation operator (recall that $\hat{S}$ is antiunitary). Then
\begin{equation}
    \hat{\Gamma}\hat{\Psi}_{0/\pi}^\dagger\hat{\Gamma}^{-1}=(-1)^\sigma\hat{\Psi}_{0/\pi}^\dagger,\label{eq:gammapsihat}
\end{equation}
so that the edge mode operator commutes or anticommutes with $\hat{\Gamma}$ depending on the chirality eigenvalue.

A bipartite lattice can be mapped to a spin chain with $2L$ sites by taking
\begin{align}
    &c_{A,n}^{(\dagger)}=c_{2n-1}^{(\dagger)}=\left[\prod_{j=1}^{2n-2}(-\sigma_j^z)\right]\sigma_{2n-1}^{-(+)},\\
    &c_{B,n}^{(\dagger)}=c_{2n}^{(\dagger)}=\left[\prod_{j=1}^{2n-1}(-\sigma_j^z)\right]\sigma_{2n}^{-(+)},
\end{align}
where $n\in\{1,...,L\}$. As an example, the Hamiltonians \eqref{eq:halphakk} satisfy the static version of \eqref{eq:appchiralH}. The corresponding free-fermion Hamiltonians are
\begin{equation}
    \hat{H}_\alpha=\sum_{n=1}^L\hc_{2(n+\alpha)-1}^\dagger\hc_{2n}+{\rm h.c.}.
\end{equation}
In spin language, these correspond to, for example,
\begin{align}
    &\hat{H}_0=\frac{1}{2}\sum_{n=1}^L(\sigma^x_{2n-1}\sigma^x_{2n}+\sigma^y_{2n-1}\sigma^y_{2n}),\\
    &\hat{H}_1=\frac{1}{2}\sum_{n=1}^{L-1}(\sigma^x_{2n}\sigma^x_{2n+1}+\sigma^y_{2n}\sigma^y_{2n+1}),\\
    &\hat{H}_{-1}=\frac{1}{2}\sum_{n=2}^{L}(\sigma^x_{2n-3}\sigma^z_{2n-2}\sigma^z_{2n-1}\sigma^x_{2n}\nonumber\\
    &\hspace{3cm}+\sigma^y_{2n-3}\sigma^z_{2n-2}\sigma^z_{2n-1}\sigma^y_{2n}),\\
    &\hat{H}_{2}=\frac{1}{2}\sum_{n=1}^{L-2}(\sigma^x_{2n}\sigma^z_{2n+1}\sigma^z_{2n+2}\sigma^x_{2n+3}\nonumber\\
    &\hspace{3cm}+\sigma^y_{2n}\sigma^z_{2n+1}\sigma^z_{2n+2}\sigma^y_{2n+3}).
\end{align}

\subsection{BDI class}

The single-particle Hamiltonians can be mapped to the BDI class by considering Majorana operators $\ha_n$, with the anticommutation relations
\begin{equation}
    \{\ha_n,\ha_m\}=2\delta_{nm}.
\end{equation}
Chiral symmetry is again given by an anti-unitary operator $\hat{S}$, where now
\begin{align}
    &\hat{S} \ha_n \hat{S}^{-1} = \sum_m\Gamma_{n,m}\ha_m, &\Gamma^2 = \Gamma^\dagger\Gamma = \id_N,\label{eq:SdefGammaMaj}
\end{align}
and, in the time-dependent case, chiral symmetry corresponds to
\begin{equation}
    \hat{S}\hat{H}(t)\hat{S}^{-1} = \Hat{H}(-t).\label{eq:SchiralHMaj}
\end{equation}
We write a quadratic Hamiltonian in the form
\begin{equation}
    \hat{H} = \frac{1}{4}\sum_{n,m}\ha_n H^M_{n,m}\ha_m,
\end{equation}
where the Majorana single-particle Hamiltonian $H^M$ is imaginary and antisymmetric. Then chiral symmetry reduces to the condition \eqref{eq:appchiralH} for $H^M$. Given a zero- or $\pi-$ mode of the single-particle Hamiltonian $H^M$, \eqref{eq:single-part}, we define the corresponding edge mode operator
\begin{align}
    \hat{\Psi}_{0/\pi} = \sum_n\psi_{0/\pi,n}\ha_n,
\end{align}
which can be shown to satisfy equations \eqref{eq:psi0piU}, \eqref{eq:gammapsihat}. Finally, one can use the imaginary condition of the Majorana single-particle Hamiltonian to show that the wavefunctions are real or come in conjugate pairs so that, for each root of $\eta$ (resp. $\zeta$), one can construct one Hermitian zero- (resp. $\pi$-) mode per boundary. As a consequence, the correlation function
\begin{equation}
    A_\infty(t) = \frac{1}{2^L}\Tr[\hat{\Psi}_{0/\pi}(\tau+t)\hat{\Psi}_{0/\pi}(\tau)]
\end{equation}
has a plateau at stroboscopic times
\begin{align}
    A_\infty(mT) = (\pm 1)^m,
\end{align}
where we used the normalization ${\hat{\Psi}_{0/\pi}}^2=1 $.

As in the previous section, one can use the single-particle Hamiltonians $H_\alpha$ to construct examples. However, note that \eqref{eq:HalphaTalpha} are not imaginary and anti-symmetric, but real and symmetric. A good way to proceed is the following. Let us diagonalize $\Gamma$ as
\begin{equation}
    \Gamma=\begin{pmatrix}
        \id_M & 0\\
        0 & -\id_M
    \end{pmatrix},
\end{equation}
where we took the $\Tr(\Gamma)=0$ for simplicity. Then
\begin{equation}
    \Gamma^{1/2}=\begin{pmatrix}
        \id_M & 0\\
        0 & i\id_M
    \end{pmatrix}
\end{equation}
and $\Gamma^{-1/2}=(\Gamma^{1/2})^{-1}$. We can define
\begin{equation}
    H^M=\Gamma^{1/2}H\Gamma^{-1/2}.\label{eq:HMgammahalf}
\end{equation}
Then $H$ is real, symmetric, and satisfies \eqref{eq:appchiralH}. In particular, we can construct
\begin{align}
    \hat{H}_\alpha&=\frac{1}{4}\sum_{n,m}\ha_n(\Gamma^{1/2}H_\alpha\Gamma^{-1/2})_{n,m}\ha_m\\
    &=\frac{i}{2}\sum_n\ha_{B,n}\ha_{A,n+\alpha}.\label{eq:HalphaMaj}
\end{align}
Finally, we note that the bipartite Majorana chain with $L$ unit cells ($2L$ sites) can be mapped to a spin-chain with $L$ sites by taking
\begin{align}
    &\ha_{A,n} = \left(\prod_{j=1}^{n-1}\sigma^z_j\right)\sigma^x_n, &\ha_{B,n} = \left(\prod_{j=1}^{n-1}\sigma^z_j\right)\sigma^y_n.
\end{align}

In particular, the unitary chiral symmetry operator action \eqref{eq:unitarySG}, given by
\begin{align}
    &\hat{\Gamma} \ha_{A,n}\hat{\Gamma} = \ha_{A,n}, &\hat{\Gamma} \ha_{B,n}\hat{\Gamma} = -\ha_{B,n},
\end{align}
can be expressed by
\begin{equation}
    \hat{\Gamma} = i^{\frac{(L-1)(L-2)}{2}}\begin{cases}
			\prod_{n=1}^L \ha_{B,n}, & \text{$L$ even}\\
            \prod_{n=1}^L \ha_{A,n}, & \text{$L$ odd}
		 \end{cases},
\end{equation}
which in the spin language corresponds to
\begin{equation}
    \hat{\Gamma} = \sigma_1^x\sigma_2^y\sigma_3^x...\label{eq:gammasigma}
\end{equation}
The Hamiltonians \eqref{eq:HalphaMaj} become
\begin{equation}
    \begin{matrix*}[l]
        &\hat{H}_0 = \frac{1}{2}\sum_n\sigma^z_n, &\hat{H}_1 = -\frac{1}{2}\sum_n\sigma^x_n\sigma^x_{n+1}\\
    &H_{-1}=-\frac{1}{2}\sum_n\sigma^y_n\sigma^y_{n+1}, &H_{2}=-\frac{1}{2}\sum_{n}\sigma^x_n\sigma^z_{n+1}\sigma^x_{n+2}.
    \end{matrix*}\nonumber
\end{equation}
Note that these Hamiltonians anticommute with \eqref{eq:gammasigma}, which agrees with the time-independent chiral symmetry, and that the amplitudes of the zero and $\pi$-modes obtained from $H$ have to be multiplied by $\Gamma^{1/2}$ as follows from equation \eqref{eq:HMgammahalf}.

\section{Numerical results in the 0-$\pi$ phase}\label{app:Numerics}

\begin{figure*}
    \includegraphics[width=0.24\textwidth]{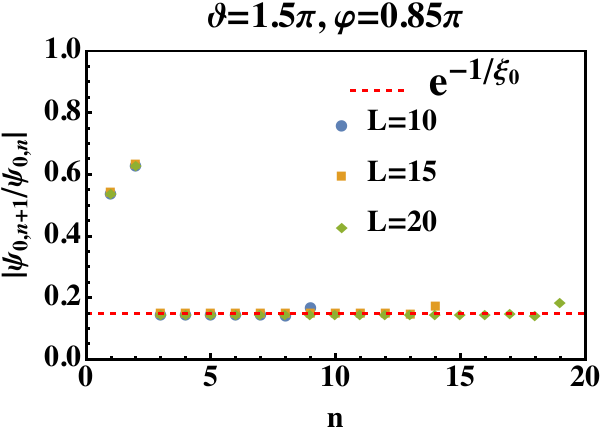}
    \includegraphics[width=0.24\textwidth]{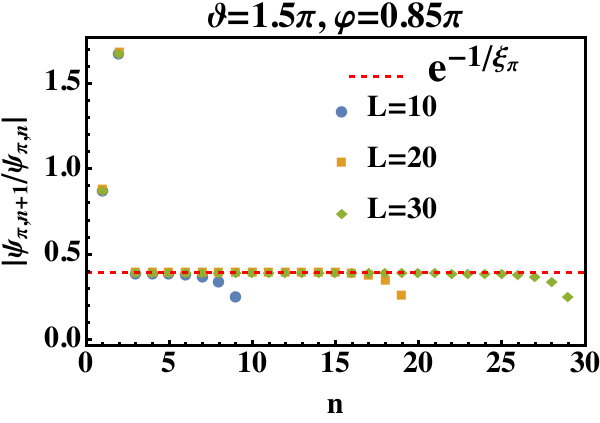}
    \includegraphics[width=0.24\textwidth]{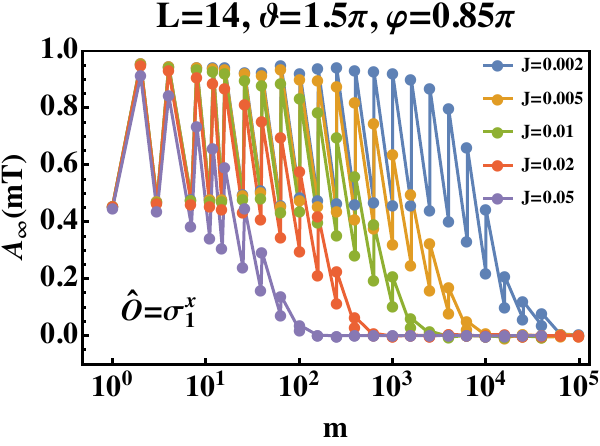}
    \includegraphics[width=0.24\textwidth]{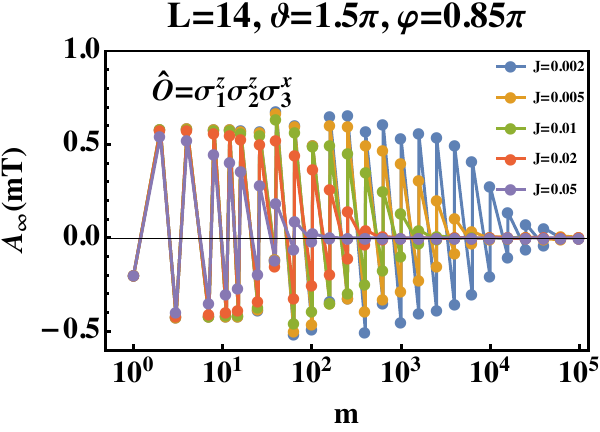}

    \caption{Left two panels: Amplitude ratio of the zero- and $\pi$-mode operators on different Majoranas along the chain obtained from ED in the Majorana basis and for different chain lengths $L$. In the bulk of the Majorana chain, the ratio agrees with the prediction $e^{-1/\xi_{0,\pi}}$ but behaves differently on the edges. The amplitude ratio for the zero-mode is always smaller than $1$, implying the maximal amplitude at the first Majorana. However, the amplitude ratio for the $\pi$-mode is greater than one elsewhere. For this example, the $\pi$-mode has maximal amplitude at the fifth Majorana.  The amplitude ratio results for zero-mode are presented only up to $L=20$ due to numerical error in computing ratios of small numbers. Right two panels: The ED results of the autocorrelation for $\sigma_1^x$ (the first Majorana) and $\sigma_1^z\sigma_2^z\sigma_3^x$ (the fifth Majorana) in the presence of perturbation \eqref{eq:perturbation} and for system size $L=14$. The autocorrelation for $\sigma_1^x$ shows features of both $0$ and $\pi$-mode dynamics and the autocorrelation for $\sigma_1^z\sigma_2^z\sigma_3^x$ mainly shows $\pi$-mode dynamics.}
    \label{fig: a1a5comparison}
\end{figure*}

\begin{figure*}
    \includegraphics[width=0.24\textwidth]{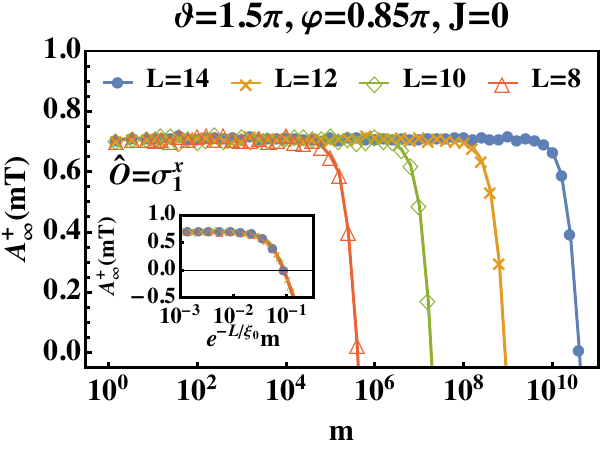}
    \includegraphics[width=0.24\textwidth]{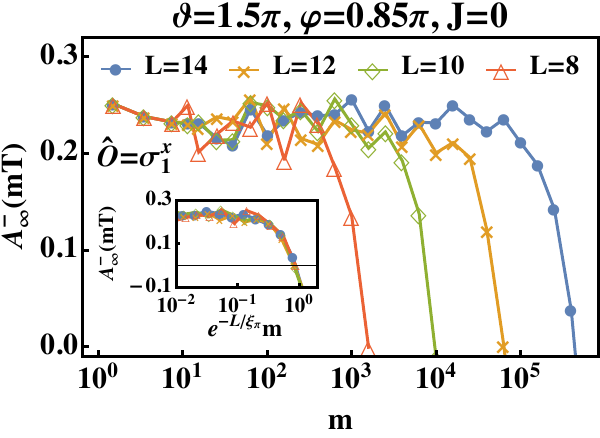}
    \includegraphics[width=0.24\textwidth]{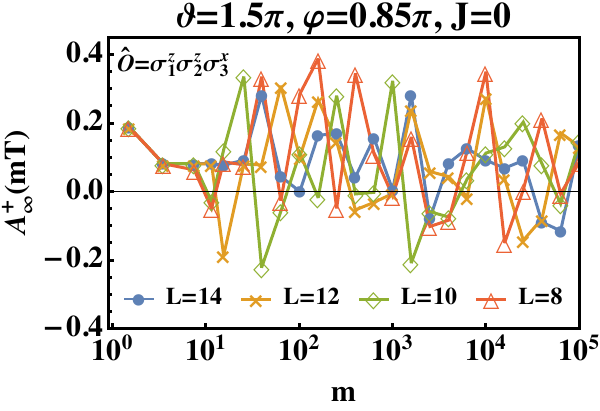}
    \includegraphics[width=0.24\textwidth]{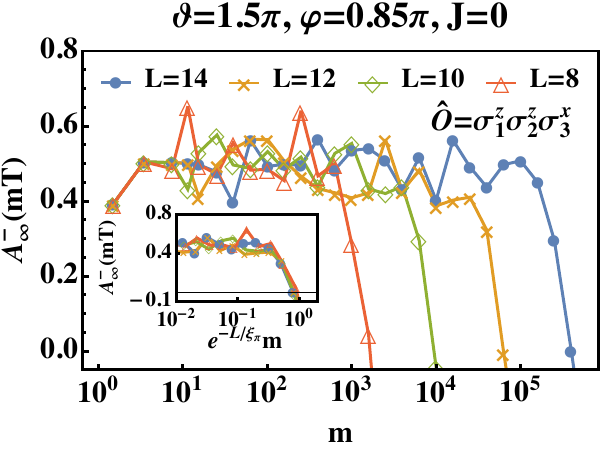}
    \includegraphics[width=0.24\textwidth]{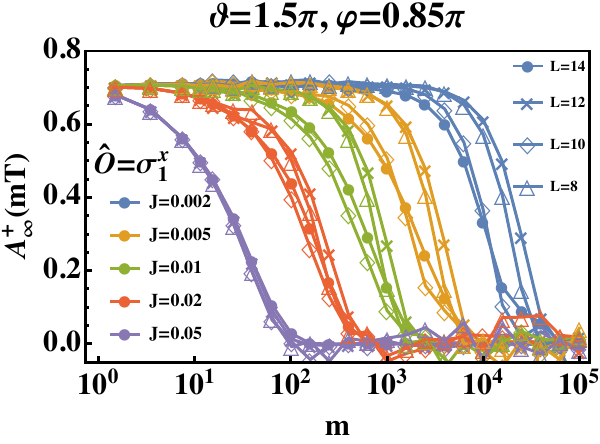}
    \includegraphics[width=0.24\textwidth]{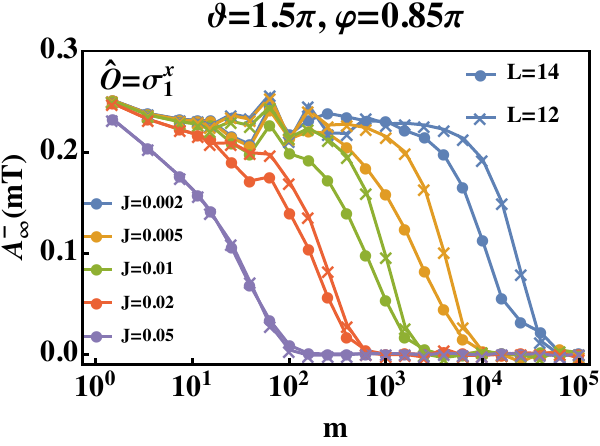}
    \includegraphics[width=0.24\textwidth]{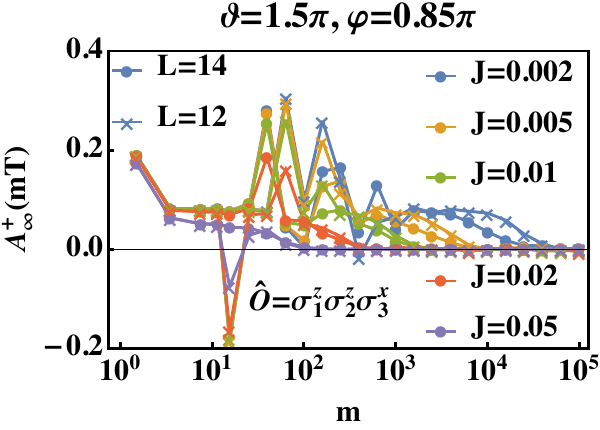}
    \includegraphics[width=0.24\textwidth]{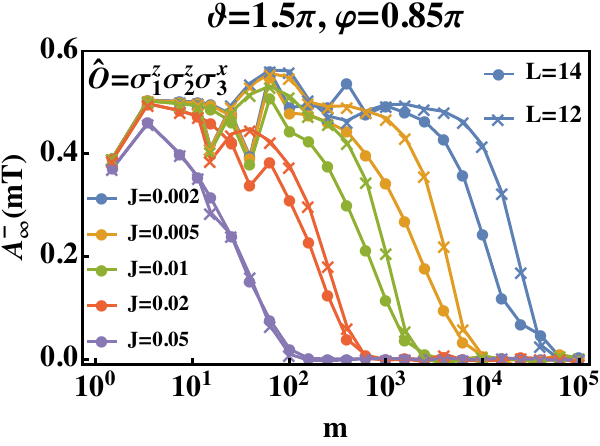}

    \caption{The ED results for the autocorrelation $A_\infty^{+}$ $\eqref{Eq: A+}$ and $A_\infty^{-}$ $\eqref{Eq: A-}$ for $\sigma^x_1$ (first two columns) and $\sigma_1^z\sigma_2^z\sigma_3^x$ (second two columns), under time evolution given by \eqref{eq:perturbation}. The parameters are $\vartheta = 1.5\pi$ and $\varphi=0.85\pi$, for which both a zero- and a $\pi$-mode exist in the non-interacting limit $J=0$. Due to the small overlap of $\sigma_1^z\sigma_2^z\sigma_3^x$ with the zero-mode, the results in the third column are presented for completeness but not used to extract the zero-mode dynamics. Top panels: System size effect in the non-interacting limit. The inset shows that the results for different system sizes $L = 8, 10, 12 ,14$ collapse when time is rescaled as $e^{-L/\xi_\pi}m$. Hence, the lifetimes of both $0$ and $\pi$-modes are exponential in system size. Bottom panel: Autocorrelations with non-zero perturbation $J$. In contrast to Fig.~\ref{Fig: autocorrelation}, the results show strong system size dependence. Systems sizes in the lower panel are chosen such that the interaction induced lifetimes are shorter than the single-particle lifetime.} 
    \label{fig: A+A-}
\end{figure*}

\begin{figure*}
    \centering
    \includegraphics[width=0.35\textwidth]{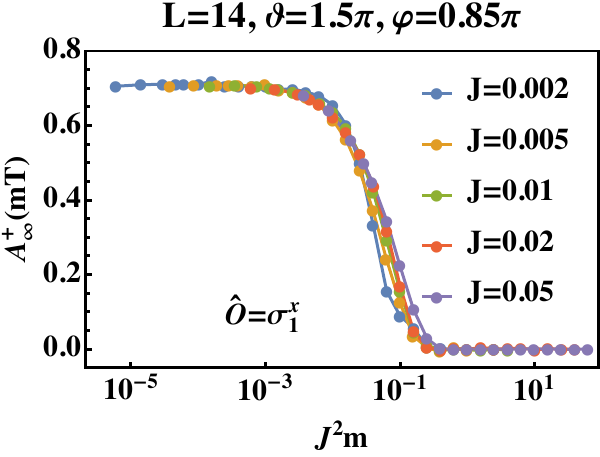}
    \includegraphics[width=0.35\textwidth]{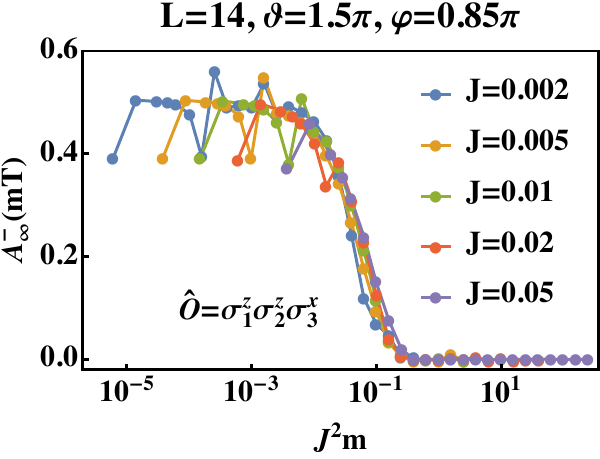}

    \caption{The rescaled autocorrelation for $L=14$ from the lower-leftmost and lower-rightmost panels in Fig.~\ref{fig: A+A-}. The data collapses at small values of the perturbation $J$ indicating a decay rate that is second order in $J$.}
    \label{fig: A+A- rescaled}
\end{figure*}

In contrast to the example in Fig.~\ref{Fig: autocorrelation}, in this section, we discuss a different case with $\vartheta = 1.5\pi$ and $\varphi = 0.85\pi$, where both a zero- and a $\pi$-mode exist in the non-interacting limit. Note that the $\pi$-mode has the same localization length as Fig.~\ref{Fig: autocorrelation} since the localization length of the $\pi$-mode is symmetric with respect to $\varphi = 1.25\pi$ on the line $\vartheta = 1.5\pi$ according to Fig.~\ref{fig:localization}. The zero-mode is peaked at the first Majorana but the $\pi$-mode has maximal peak at the fifth Majorana as shown in the left two panels of Fig.~\ref{fig: a1a5comparison}. Therefore, we compute the autocorrelation functions of the first ($\sigma_1^x$) and of the fifth Majoranas ($\sigma_1^z\sigma_2^z\sigma_3^x$) and then extract the zero- and $\pi$-mode dynamics from them separately. The ED results with perturbation at system size $L=14$ are presented in the two right panels of Fig.~\ref{fig: a1a5comparison}. The autocorrelation of $\sigma_1^x$ shows both zero- and $\pi$-mode dynamics, while the autocorrelation of $\sigma_1^z\sigma_2^z\sigma_3^x$ mainly shows $\pi$-mode dynamics. The decays of the zero- and $\pi$-mode can be studied directly through the following decomposition
\begin{align}
    &A_\infty^{+}(mT+T/2) = \frac{A_\infty (mT+T)+A_\infty (mT)}{2},\label{Eq: A+}\\
    &A_\infty^{-}(mT+T/2) = \frac{A_\infty (mT+T) -A_\infty (mT)}{2}.\label{Eq: A-}
\end{align} 
We utilize the data points where $m$ is odd in \eqref{Eq: A+} and \eqref{Eq: A-}, ensuring a positive sign for $A^{-}_{\infty}$.

The corresponding $A^{+}_{\infty}$ and $A^{-}_{\infty}$ results are presented in Fig.~\ref{fig: A+A-}. Similar to the top panel in Fig.~\ref{Fig: autocorrelation}, the plots in the first row of Fig.~\ref{fig: A+A-} shows that the lifetimes of the zero- and $\pi$-modes are exponential in system size $L$ in the non-interacting limit. For $A^{+}_{\infty}$ of $\sigma_1^z\sigma_2^z\sigma_3^x$, the overlap with the zero-mode is too small to be used to extract the zero-mode dynamics. The second row in Fig.~\ref{fig: A+A-} shows the effects of interactions. One distinct feature in Fig.~\ref{fig: A+A-} is the strong system size dependence of the autocorrelation, e.g., in the leftmost plot of the second row. For small values of the perturbation strength, the $L = 8, 12$ results have longer lifetime than the $L = 10, 14$ results. The autocorrelation does not converge with increasing system size but seems to oscillate between two different limits. A surprising observation is that the single-particle lifetime is longer than $10^5$ for system sizes $L=8, 10, 12, 14$ (the leftmost plot of $A^{+}_{\infty}$ in the top row), and yet once interactions are switched on, the dynamics shows strong system size dependence even for relatively large perturbations (see lower left-most panel). A possible explanation for this effect could be the spatial distance between the zero-mode (peaked at the first Majorana) and the $\pi$-mode (peaked at the fifth Majorana), leading to an additional system size dependence in the autocorrelation. Further studies are needed to clarify this observation.

The $\pi$ mode in Fig.~\ref{fig: A+A-} has a zero-mode partner unlike the $\pi$ mode in Fig.~\ref{Fig: autocorrelation}. Thus when interactions are switched on, 
the first $\pi$ mode has an additional scattering channel involving scattering with the zero mode, making its lifetime shorter than that of the $\pi$ mode in Fig.~\ref{Fig: autocorrelation}. This idea is supported by the rescaled autocorrelations in Fig.~\ref{fig: A+A- rescaled}. The top (bottom) panel shows that the $L=14$ rescaled autocorrelation $A^{+}_{\infty} (A^{-}_{\infty})$ for $\sigma_1^x (\sigma_1^z\sigma_2^z\sigma_3^x)$ is consistent with a Fermi's Golden Rule decay rate for the zero-mode ($\pi$-mode). Note that both the zero-mode and the $\pi$-mode have similar decay rates despite their localization lengths being different (left two panels in Fig.~\ref{fig: a1a5comparison}). This indicates that the decay rates are dominated by the scattering between the zero- and the $\pi$-modes rather than the scattering of these modes independently with the bulk excitations \cite{yeh2023decay,YehProduct}.

\bibliographystyle{apsrev4-1}
\bibliography{floquet}

\end{document}